\documentclass[
floatfix,
twocolumn,
showpacs,showkeys,preprintnumbers,aps,prd]
{revtex4}

\usepackage{amsmath}
\usepackage{amssymb}
\usepackage{revsymb}
\usepackage{graphicx}
\usepackage{dcolumn}

\begin{document}

\title{Electromagnetic nucleon form factors in instant and point form}

\author{T. Melde$^1$, K. Berger$^1$, L. Canton$^2$, W. Plessas$^1$ and R. F. Wagenbrunn$^1$}

\address{$^1$ Theoretische Physik, Institut f\"ur Physik,
Karl-Franzens-Universit\"at Graz,
Universit\"atsplatz 5, A-8010 Graz, Austria}
\address{$^2$ Istituto Nazionale di Fisica Nucleare, Sezione di Padova, 
Via F. Marzolo 8, I-35131 Padova, Italy}
%
\begin{abstract}
We present a study of the electromagnetic structure of the nucleons with
constituent quark models in the framework of relativistic quantum mechanics.
In particular, we address the construction of spectator-model currents in
the instant and point forms. Corresponding results for the elastic nucleon
electromagnetic form factors as well as charge radii and magnetic moments
are presented. We also compare results obtained by different realistic nucleon 
wave functions stemming from alternative constituent quark models. 
Finally, we discuss the theoretical uncertainties that reside in the construction 
of spectator-model transition operators.
\end{abstract}
\pacs{12.39.Ki,13.40.-f,11.30.Cp,11.30.Er}
\vspace{2pc}
\keywords{
Electromagnetic nucleon form factors; Electric radii; Magnetic moments;
Constituent quark model; 
Relativistic quantum mechanics; Instant form; Point form
}
\maketitle
\section{Introduction}
A promising approach to the nucleon electromagnetic (EM) structure at low and moderate
momentum transfers consists in employing constituent quark models (CQMs) in the
framework of relativistic quantum mechanics (RQM). Different forms of RQM,
such as the instant, front, and point forms,
have been used by various authors (e.g., in 
Refs.~\cite{Cardarelli:1995dc,Szczepaniak:1995mi,Merten:2002nz,Pace:2003aa,Julia-Diaz:2003gq}).
Realistic descriptions of the EM nucleon form factors were specifically obtained
with the Goldstone-boson-exchange (GBE) CQM. Working within the point form,
the electric and magnetic form factors of
both the proton and the neutron were described in surprisingly good agreement
with experimental data~\cite{Wagenbrunn:2000es,Boffi:2001zb}. Similarly, the electric
radii and magnetic moments of all the octet and decuplet baryon ground states
were reproduced as well~\cite{Berger:2004yi}. By an analogous
calculation also the axial nucleon form factors could be explained
consistently~\cite{Glozman:2001zc}. Until now, it remains as a puzzle why
the direct quark-model predictions in point form fall so close to experiment
in all aspects of the electroweak nucleon structure, especially since simplified
spectator-model current operators have been employed and no additional
parametrizations have been introduced. It is
noteworthy that the point-form results are also very similar to the
parameter-free predictions of the instanton-induced (II) CQM of the Bonn
group that were calculated in a completely different approach along the
Bethe-Salpeter formalism~\cite{Merten:2002nz}.
The quality of the reproduction of the baryon 
EM properties gave impulse to studies of strong decays of baryon resonances
along the same formalisms. However,
the corresponding decay widths are not described equally
well~\cite{Melde:2005hy,Melde:2004qu,Sengl:2006gz,Melde:2006yw,Sengl:2007yq,Metsch:2003ix}. 

In order to clarify the situation we undertook 
a closer inspection of the point-form spectator model (PFSM) in
Ref.~\cite{Melde:2004qu}. We demonstrated along the strong decay widths
that some significant effects may be caused by ambiguities connected to the
fact that a unique spectator-model construction cannot be obtained by
imposing Poincar\'e invariance alone. It was already observed before that
further constraints should  be
included~\cite{Polyzou:1986df,Lev:1995ann}. This property is inherent to any
form of RQM and in this paper we specifically discuss the effects of such ambiguities 
in the instant form spectator model (IFSM) and the PFSM.

We address the proton and neutron electromagnetic form factors at momentum
transfers up to $Q^2 = 4$ GeV$^2$. First we introduce the construction of
instant-form and point-form spectator-model currents and characterize their
particular properties. We then derive their nonrelativistic limits and show
that both, IFSM and PFSM, lead to the same result, namely, the well-known 
nonrelativistic impulse approximation (NRIA). We provide a comparison of the 
IFSM, PFSM and NRIA predictions of the GBE
CQM~\cite{Glozman:1998ag}. 
Subsequently, we examine the effects from different realistic CQMs relying on distinct
quark-quark dynamics. In particular, we discuss the PFSM results by the GBE
CQM in relation to the ones of a
one-gluon-exchange (OGE) CQM, namely the relativistic version of the
Bhaduri-Cohler-Nogami (BCN) CQM~\cite{Bhaduri:1981pn} as parametrized in
Ref.~\cite{Theussl:2000sj}. In this context a comparison is made also
with the form-factor results reported from the II CQM by the Bonn
group~\cite{Merten:2002nz}. Finally, in line with 
Refs.~\cite{Polyzou:1986df,Lev:1995ann}, we
discuss how additional constraints like charge normalization and time-reversal
invariance can be exploited to reduce as much as possible some remaining
ambiguities in the present form of the spectator-model constructions. We end
the paper in Sect.~\ref{sec:Conclusion} with a summary and conclusion.
\section{Electromagnetic Current}
\label{sec:2}
In RQM the nucleon states are expressed as eigenstates $\left|P,J,\Sigma \right>$
of the interacting mass operator
\begin{equation}
\label{eq:masses}
\hat M = \hat M_{\rm free} + \hat M_{\rm int} \, ,
\end{equation}
the intrinsic spin operator $\hat J$, and its z-component $\hat \Sigma$ (the
letters without hat denoting the corresponding eigenvalues). The covariant normalization
of these states is
\begin{equation}
\left<P', J',\Sigma'|P, J,\Sigma\right>=
2P_0\delta^3\left({\vec P}-{\vec P}'\right)\delta_{JJ'}
\delta_{\Sigma\Sigma'} \, .
\label{normP}
\end{equation}
The mass operator $\hat M$ is connected to the four-momentum operator $\hat P$ and
the four-velocity operator $\hat V$ by the relations
\begin{equation}
    \hat P^{\mu}={\hat M}\hat V^{\mu} \, , 
\end{equation}
where
\begin{equation}
\hat P^{\mu} \hat P_{\mu}=\hat M^2 \, .
\end{equation}
Because of the commutation relations among these operators the nucleon eigenstates
with mass $M$
can equivalently be denoted as $\left|V,M,J,\Sigma\right>$. In this notation the
elastic transition amplitude between the incoming and outgoing nucleon eigenstates 
is given by the following matrix element of the reduced electromagnetic current operator
\begin{equation}
F^{\mu}_{\Sigma',\Sigma}\left(Q^2\right)
=
\left<V',M,\frac{1}{2},\Sigma'\right|{\hat J}^{\mu}_{\rm rd}
\left|V,M,\frac{1}{2},\Sigma\right>
\label{eq:Fmu}\, ,
\end{equation}
where $Q^2=(P'^\mu-P^\mu)^2$ is the momentum transfer by the virtual photon.
In the Breit frame, the nucleon electric and magnetic Sachs form factors,
$G_{\rm E}\left(Q^2\right)$ and $G_{\rm M}\left(Q^2\right)$,
are related to the transition amplitude 
$F^{\mu}_{\Sigma',\Sigma}$ by
\begin{eqnarray}
\label{ge}
F^{0}_{\Sigma',\Sigma}\left(Q^2\right)
&=&2M G_{\rm E}\left(Q^2\right)\delta_{\Sigma'\Sigma}
\\
\label{gm}
{\vec F}_{\Sigma',\Sigma}\left(Q^2\right)
&=&
iQ G_{\rm M}\left(Q^2\right)
\chi^\dagger_{\Sigma'}\left({\vec \sigma} \times {\vec e}_{z}\right)\chi_\Sigma^{\phantom\dagger}
\, ,
\end{eqnarray}
where $\Sigma', \Sigma=\pm \frac{1}{2}$ are the projections of the nucleon spin
$\vec \sigma$ along the direction ${\vec e}_{z}$ of the
$z$-axis and $\chi$ are the corresponding Pauli spinors. The form factors also
lead directly to the magnetic moments
\begin{equation}
\mu = G_{\rm M}\left(Q^2=0\right)
\end{equation}
and the charge radii
\begin{equation}
r^2_{\rm ch}=\left.-6\frac{dG_{\rm E}}{d\left(Q^2\right)}\right|_{Q^2=0}\, .
\end{equation}

The electric and magnetic form factors in Eqs.~(\ref{ge}) and~(\ref{gm}) are
Poincar\'e invariant, since both the mass-operator eigenstates and the
electromagnetic current operator transform under the Poincar\'e
group. Depending on the particular form of RQM (instant, front, or point
forms) certain generators of the
Poincar\'e transformations become interaction dependent while the other ones
belong to the kinematical subgroup.

It is still rather difficult to employ the full (many-body) structure of the
current operator and thus one adheres to simplifications. The common form 
consists in the so-called spectator approximation. 
The definition of a spectator-model current operator is generally not 
unique and requires additional constraints. In particular, the
spectator-model construction has to be covariant under the transformations
of the kinematic subgroup of the particular form of RQM and it must guarantee
for time-reversal invariance. Further it should
reduce to a sum of genuine single-particle currents in the limit of vanishing
interaction among the constituent quarks and yield the charge of the
proton for the electric form factor approaching
momentum transfer $Q^2=0$. In addition, the construction should lead to a proper
nonrelativistic limit.
\subsection{Instant-Form Spectator Model}
In the instant form the spatial translations and rotations form the kinematic
subgroup, whereas the boosts become interaction dependent. 

In the explicit calculations of the matrix elements~(\ref{eq:Fmu}) one requires
momentum eigenstates of the free three-quark system
$\left|p_1,p_2,p_3;\sigma_1,\sigma_2,\sigma_3\right>$ defined as tensor products
of single-particle momentum eigenstates $\left|p_i;\sigma_i\right>$. 
In any reference frame with total three-momentum $\vec P=\sum_i{\vec p_i}$
they can also be expressed as 
\begin{multline}
\label{eq:ifP}
\left|\vec P; \vec k_1,\vec k_2,\vec k_3; \mu_1,\mu_2,\mu_3\right>=\\
\sum_{\sigma_i }\prod_{i}{D_{\sigma_i\mu_i}^{\frac{1}{2}}
\left\{R_W\left[k_i;B\left(v\right)\right]\right\}
}
\left|p_1,p_2,p_3;\sigma_1,\sigma_2,\sigma_3\right>\, ,
\end{multline}
wherein the Wigner rotations depend on the free four velocity
\begin{equation}
v=\frac{P_{\rm free}}{M_{\rm free}}=\frac{\sum_i{p_i}}{\sum_i{\omega_i}} \, .
\end{equation}
The momenta $k_i$ are connected to the momenta $p_i$ by the Boost relations
$p_i=B(v)k_i$ and they fulfill the constraint $\sum_i{\vec k_i}=0$.
The individual quark energies are given by $\omega_i=\sqrt{m_i^2+\vec k_i^2}$.
The $\mu_i$ correspond to the individual quark spins in the rest frame of
the three-quark system. The free momentum eigenstates of Eq.~(\ref{eq:ifP})
have the following completeness relation
\begin{multline}
\mathbf{1}=\sum_{\mu_i}{\int{d^3\vec P d^3\vec k_2 d^3\vec k_3
\frac{\sum_i{\omega_i}}{E_{\rm free}}
\frac{1}{2\omega_1 2\omega_2 2\omega_3}
}}\\
\times \left|\vec P; \vec k_1,\vec k_2,\vec k_3; \mu_1,\mu_2,\mu_3\right>
\left<\vec P; \vec k_1,\vec k_2,\vec k_3; \mu_1,\mu_2,\mu_3\right|
\, .
\end{multline}
They are well suited for instant-form calculations because here the three-momentum
$\vec P$ is not affected by interactions. Representing the nucleon states with
this basis allows to separate the internal motion according to
\begin{multline}
\label{eq:wfuncif}
\left<\vec P'; \vec k'_1,\vec k'_2,\vec k'_3; \mu'_1,\mu'_2,\mu'_3|\vec P,J, \Sigma\right>=\\
\sqrt{2E'_{\rm free}E}\sqrt{\frac{2\omega'_1 2\omega'_2 2\omega'_3}{\sum{\omega'_i}}}
\delta^3\left(\vec P'-\vec P\right)\Psi_{M\frac{1}{2}\Sigma}\left({\vec k}_i;\mu_i\right)\, ,
\end{multline}
where $E'_{\rm free}$ and $E$ are the eigenvalues of the zeroth components of the free
and interacting four-momenta $\hat P'^\mu_{\rm free}$ and $\hat P^\mu$, respectively.
The wave function $\Psi_{M\frac{1}{2}\Sigma}\left({\vec k}_i;\mu_i\right)$ is just
the rest-frame wave function of the nucleon. It is normalized to unity
\begin{equation}
\label{eq:wavenorm}
  \sum_{\mu_{i}}
    \int{d^{3}k_{2}d^{3}k_{3}}
   \Psi^{\star}_{M\frac{1}{2}\Sigma'}\left( 
    \vec{k}_i;\mu_i
   \right)
   \Psi_{M\frac{1}{2}\Sigma}\left(
   \vec{k}_i;\mu_i
   \right) = \delta_{\Sigma\Sigma'}
\end{equation}
in accordance with the normalization condition of the nucleon eigenstates in
Eq.~(\ref{normP}).
 
The transition amplitude~(\ref{eq:Fmu}) in instant form can then be expressed as
\begin{widetext}
\begin{eqnarray}
F^{\mu}_{\Sigma',\Sigma}\left(Q^2\right)
&=&2\sqrt{EE'}
\sum_{\sigma_i\sigma'_i}\sum_{\mu_i\mu'_i}{
\int{
d^3{\vec k}_2d^3{\vec k}_3d^3{\vec k}'_2d^3{\vec k}'_3
}}
\frac{1}{\sqrt{E_{\rm free}E'_{\rm free}}}
\sqrt{\frac{\sum{\omega_i}}{2\omega_1 2\omega_2 2\omega_3}}
\sqrt{\frac{\sum{\omega'_i}}{2\omega'_1 2\omega'_2 2\omega'_3}}
\nonumber\\
&&
{
\Psi^\star_{M\frac{1}{2}\Sigma'}\left({\vec k}'_i;\mu'_i\right)
\prod_{\sigma'_i}{D_{\sigma'_i\mu'_i}^{\star \frac{1}{2}}
\left\{R_W\left[k'_i;B\left(v'\right)\right]\right\}
}}
\nonumber\\
&&
\left<p'_1,p'_2,p'_3;\sigma'_1,\sigma'_2,\sigma'_3\right|
{\hat J}^{\mu}_{\rm rd}
\left|p_1,p_2,p_3;\sigma_1,\sigma_2,\sigma_3\right>
\prod_{\sigma_i}{D_{\sigma_i\mu_i}^{\frac{1}{2}}
\left\{R_W\left[k_i;B\left(v\right)\right]\right\}
}
\Psi_{M\frac{1}{2}\Sigma}\left({\vec k}_i;\mu_i\right)
\label{eq:Fmuif}\, ,
\end{eqnarray}
\end{widetext}
The spectator model of the current operator in instant form
(the IFSM) is then defined as
\begin{multline}
\label{eq:ifsm}
\left<p'_1,p'_2,p'_3;\sigma'_1,\sigma'_2,\sigma'_3\right|
{\hat J}^{\mu}_{\rm rd,IFSM}
\left|p_1,p_2,p_3;\sigma_1,\sigma_2,\sigma_3\right>
=
\\
3e_1{\bar u}\left(p'_1,\sigma'_1\right)
\gamma^\mu
u\left(p_1,\sigma_1\right) 
\\
2p_{20}\delta^3\left({\vec p}_2-{\vec p}'_2\right)
2p_{30}\delta^3\left({\vec p}_3-{\vec p}'_3\right)
 \delta_{\sigma_{2}\sigma'_{2}}
   \delta_{\sigma_{3}\sigma'_{3}}
   \, .
\end{multline}
As a consequence of the very properties of the instant form (with the
three-momenta as generators of spatial translations lying in the kinematic
subgroup) the momenta of the struck quark in the incoming and outgoing nucleon
are related by
\begin{eqnarray}
\label{eq:p1ifsm}
p_{10}-p'_{10}&=&\tilde q_0 \nonumber \\ 
\vec p_1-\vec p'_1&=&\vec Q \, .
\end{eqnarray}
This means that the whole three-momentum carried by the virtual photon is
transferred to the quark 1, while only a part of the photon energy is absorbed
by a single quark, i.e. $\tilde q_0 \ne Q_0$. Clearly, $\tilde q_0$ is uniquely
determined by overall momentum conservation and the two spectator conditions.
We shall see below that in the point form the spectator-model construction leads
to a different relation between $p_1$ and $p'_1$, as the momenta no longer
lie in the kinematic subgroup. 

The construction (\ref{eq:ifsm}) is usually adopted in the Breit frame.
When transformed to a different reference frame, it does not preserve
its spectator-model structure but acquires additional many-body contributions.
Thus, in general, the IFSM current should not be viewed as a one-body operator.
Furthermore a spectator-model construction made in another reference frame
defines a different spectator-model current. As a result the calculation done
with an IFSM current defined in the Breit frame leads to a different result
than the calculation performed with an IFSM current defined in the laboratory
frame, say. In this sense the IFSM, while always yielding Poincar\'e invariant
results, bears an ambiguity in the construction, as it is per se frame dependent.
If one imposes time-reversal invariance on the IFSM, one necessarily has to
resort to the Breit frame. In all other reference frames additional many-body
contributions would be needed to guarantee for time-reversal invariance.

Eq.~(\ref{eq:Fmuif}) exhibits all the effects of the Lorentz boosts on the
incoming and outgoing nucleon states through the changes in the respective
quark momenta and the Wigner $D$-functions. Sometimes the latter are simply
ignored by setting them to unity~\cite{Li:2007hn}. While this simplifies the calculations
considerably, the resulting form factors obtained in this way
are no longer strictly Poincar\'e invariant. 
\subsection{Point-Form Spectator Model}
In the point form the kinematic subgroup is the Lorentz group, only the space-time
translations are interaction dependent. For the actual calculations it is advantageous
to use velocity states (of the free three-quark system) defined by
\begin{multline}
\label{eq:ifV}
\left|v; \vec k_1,\vec k_2,\vec k_3; \mu_1,\mu_2,\mu_3\right>=\\
\sum_{\sigma_i }\prod_{i}{D_{\sigma_i\mu_i}^{\frac{1}{2}}
\left\{R_W\left[k_i;B\left(v\right)\right]\right\}
}
\left|p_1,p_2,p_3;\sigma_1,\sigma_2,\sigma_3\right>\, ,
\end{multline}
where the momenta $k_i$ and $p_i$ are again connected through the boost relation
$p_i=B\left(v\right)k_i$ and the $k_i$ satisfy $\sum_i{\vec k_i}=0$.

Since in the point form the four-velocity $v=(v_0,\vec v)$ is
independent of the interaction, it can be expressed through eigenvalues of the
free or interacting momentum and mass operators as
\begin{equation}
v=\frac{P_{\rm free}}{M_{\rm free}}=\frac{P}{M}=V \, .
\end{equation}
The completeness relation for the velocity states reads
\begin{multline}
{\mathbf 1}=\sum_{\mu_{i}}\int{
    d^{3}{\vec v}
    d^{3}{\vec k}_{2}
    d^{3}{\vec k}_{3}
    \frac{\left(\sum_i{\omega_{i}}\right)^{3}}{v_{0}}
    \frac{1}{2\omega_{1}2\omega_{2}2\omega_{3}}
    }\\
    \times \left|v;\vec{k}_1,\vec{k}_2,\vec{k}_3;
\mu_1,\mu_2,\mu_3\right\rangle
    \left\langle 
    v;\vec{k}_1,\vec{k}_2,\vec{k}_3;\mu_1,\mu_2,\mu_3\right|\, .
\label{eq:velcomp}
\end{multline}
Representing the nucleon states in the velocity-state basis allows to
separate the internal motion in the following way
\begin{multline}
\left\langle v;\vec{k}_1,\vec{k}_2,\vec{k}_3;\mu_1,\mu_2,\mu_3
    |
    V,M,\frac{1}{2},\Sigma\right\rangle
    \\
    =\frac{\sqrt{2}}{M} v_{0}\delta^{3}\left(\vec{v}-\vec{V}\right)
   \sqrt{\frac{2\omega_{1}2\omega_{2}2\omega_{3}}
    {\left(\sum_i{\omega_{i}}\right)^{3}}
    }
   \Psi_{M\frac{1}{2}\Sigma}\left(
   \vec{k}_i
   ;\mu_i
   \right) \, ,
   \label{eq:wavefuncpoint}
\end{multline}
which differs substantially from the separation followed in the instant form,
Eq.~(\ref{eq:wfuncif}). However, the
$\Psi_{M\frac{1}{2}\Sigma}\left(\vec{k}_i;\mu_i\right)$ is again the
rest-frame wave function of the nucleon. The nucleon mass eigenstates
in Eq.~(\ref{eq:wavefuncpoint}) are normalized as in Eq.~(\ref{normP}) and the
wave functions are normalized as in Eq.~(\ref{eq:wavenorm}). 

The transition amplitude, Eq.~(\ref{eq:Fmu}), of the electromagnetic current in point
form reads
\begin{widetext}
\begin{eqnarray}
F^{\mu}_{\Sigma',\Sigma}\left(Q^2\right)
&=&
\frac{2}{M^2}\sum_{\sigma_i\sigma'_i}\sum_{\mu_i\mu'_i}{
\int{
d^3{\vec k}_2d^3{\vec k}_3d^3{\vec k}'_2d^3{\vec k}'_3
}}
\sqrt{\frac{\left(\omega_1+\omega_2+\omega_3\right)^3}
{2\omega_1 2\omega_2 2\omega_3}}
\sqrt{\frac{\left(\omega'_1+\omega'_2+\omega'_3\right)^3}
{2\omega'_1 2\omega'_2 2\omega'_3}}
\nonumber\\
&&
{
\Psi^\star_{M\frac{1}{2}\Sigma'}\left({\vec k}'_i;\mu'_i\right)
\prod_{\sigma'_i}{D_{\sigma'_i\mu'_i}^{\star \frac{1}{2}}
\left\{R_W\left[k'_i;B\left(V'\right)\right]\right\}
}}
\nonumber\\
&&
\left<p'_1,p'_2,p'_3;\sigma'_1,\sigma'_2,\sigma'_3\right|
{\hat J}^{\mu}_{\rm rd, PFSM}
\left|p_1,p_2,p_3;\sigma_1,\sigma_2,\sigma_3\right>
\prod_{\sigma_i}{D_{\sigma_i\mu_i}^{\frac{1}{2}}
\left\{R_W\left[k_i;B\left(V\right)\right]\right\}
}
\Psi_{M\frac{1}{2}\Sigma}\left({\vec k}_i;\mu_i\right)
\label{eq:FmuPF}\, ,
\end{eqnarray}
\end{widetext}
and the spectator-model approximation of the current operator in point form (the
PFSM) is defined by the expression
\begin{multline}
\label{eq:pfsm}
\left<p'_1,p'_2,p'_3;\sigma'_1,\sigma'_2,\sigma'_3\right|
{\hat J}^{\mu}_{\rm rd, PFSM}
\left|p_1,p_2,p_3;\sigma_1,\sigma_2,\sigma_3\right>
=
\\
3 {\cal N}
e_1{\bar u}\left(p'_1,\sigma'_1\right)
\gamma^\mu
u\left(p_1,\sigma_1\right) 
\\
2p_{20}\delta^3\left({\vec p}_2-{\vec p}'_2\right)
2p_{30}\delta^3\left({\vec p}_3-{\vec p}'_3\right)
 \delta_{\sigma_{2}\sigma'_{2}}
   \delta_{\sigma_{3}\sigma'_{3}} \, .
\end{multline}
Here the momentum transfer to the struck quark is given by
\begin{equation}
\label{eq:p1pfsm}
p^\mu_1-{p'}^\mu_1= \tilde q^\mu \, ,
\end{equation}
where $\tilde q^\mu \ne Q^\mu$ is uniquely
determined by the overall momentum conservation and the two spectator conditions.
The relation between $p_1$ and $p'_1$ is complicated, because the transferred
momentum $\tilde q$ to the active quark inherits nontrivial interaction-dependent
contributions~\cite{Melde:2004qu}. The momentum transfers to the struck quark 
in IFSM, Eq.~(\ref{eq:p1ifsm}), and in PFSM, Eq.~(\ref{eq:p1pfsm}), are rather
different, because in the instant form the kinematical subgroup includes the 
three-momentum, whereas in point form it does not.
Contrary to the IFSM, the PFSM maintains its spectator-model character in
all reference frames. Always, only one quark is directly coupled to the photon,
and the PFSM current is manifestly covariant. 
In the limit of vanishing quark-quark interaction such differences disappear 
and one recovers $\tilde q^\mu \rightarrow Q^\mu$.

Formally, the above definition of Eq.~(\ref{eq:pfsm}) looks the same as for 
the IFSM in Eq.~(\ref{eq:ifsm}) except for the appearance of an extra factor $\cal N$.  
The factor $\cal N$ is needed in the point form in order
to yield the proper proton charge from the electric form factor in the limit
$Q^2 \rightarrow 0$ and to recover a sum of individual-particle currents when
the interaction is turned off~\cite{Melde:2004qu}. In the
works~\cite{Wagenbrunn:2000es,Glozman:2001zc,Boffi:2001zb,Berger:2004yi} a choice
of $\cal N$ symmetric with regard to the incoming and outgoing nucleon states was
adopted
\begin{equation}
{\cal N}={\cal N}_{\rm S}=
\left(\frac{M}{\sum_i{\omega_{i}}}\right)^{\frac{3}{2}}
\left(\frac{M}{\sum_i{\omega'_{i}}}\right)^{\frac{3}{2}}\, .
\label{eq:geoN}
\end{equation}
It is important to notice that the factor $\cal N$ has nothing to do with the
normalization of the nucleon states: the nucleon states follow the covariant 
normalization as in Eq. (\ref{normP}) implying, together with the normalization
of the wave function in Eq.~(\ref{eq:wavenorm}), the factors explicitly written 
out in Eq. (\ref{eq:FmuPF}); the factor ${\cal N}$ must be considered as part of 
the definition of the PFSM {\it current} and it is required to produce a 
consistent spectator-model operator in the point form.

One should note that both $\cal N$ and $\tilde q$ effectively depend on all three 
quark variables, and therefore the PFSM must not be considered strictly as a
one-body operator. It effectively includes many-body
contributions~\cite{Melde:2004qu}. The factor $\cal N$
can in principle be adopted in several ways. Later on we shall come back to this
issue.

\subsection{Nonrelativistic Impulse Approximation}
The nonrelativistic reduction of both the IFSM and the PFSM leads to the same
result, namely, to the nonrelativistic impulse approximation (NRIA):
\begin{widetext}
\begin{eqnarray}
F^{\mu \rm NR}_{\Sigma',\Sigma}\left(Q^2\right)
&=&
\left<V',M,\frac{1}{2},\Sigma'\right|{\hat J}^{\mu}_{\rm rd, NR}
\left|V,M,\frac{1}{2},\Sigma\right>
=2M
\sum_{\mu_i\mu'_i}
\int{
d^3{\vec k}_2d^3{\vec k}_3d^3{\vec k}'_2d^3{\vec k}'_3
}
\nonumber\\
&&
\Psi^\star_{M\frac{1}{2}\Sigma'}\left({\vec k}'_i;\mu'_i\right)
{j}^{\mu}_{\rm rd, NR}\left(Q^2\right)
\Psi_{M\frac{1}{2}\Sigma}\left({\vec k}_i;\mu_i\right)
\label{eq:Fmunr}
\end{eqnarray}
\end{widetext}
with
\begin{multline}
{j}^{\mu}_{\rm rd, NR}\left(Q^2\right)=\\
e_1
\begin{pmatrix}
1\\ \\
\left[ 
 \frac{\vec p_1 +\vec p'_1}{2m_1}+\frac{i\vec\sigma_1\times
\left(\vec p_1-\vec p'_1\right)}{2m_1}
\right]
\delta^3\left(\vec p_2-\vec p'_2\right)
\delta^3\left(\vec p_3-\vec p'_3\right)
\end{pmatrix}
\end{multline}
and the momenta now being connected through the relations
$\vec k_i=\vec p_i-\frac{m_i}{\sum_i{m_i}}\vec P$ and
$\vec p_1-\vec p'_1=\vec Q$.

As is evident, all integration measures, normalization factors, and Wigner
rotations have reduced to unity and the nonrelativistic expression for the
EM current is recovered.

\begin{figure*}[t]
\begin{center}
\includegraphics[clip=,height=6.cm]{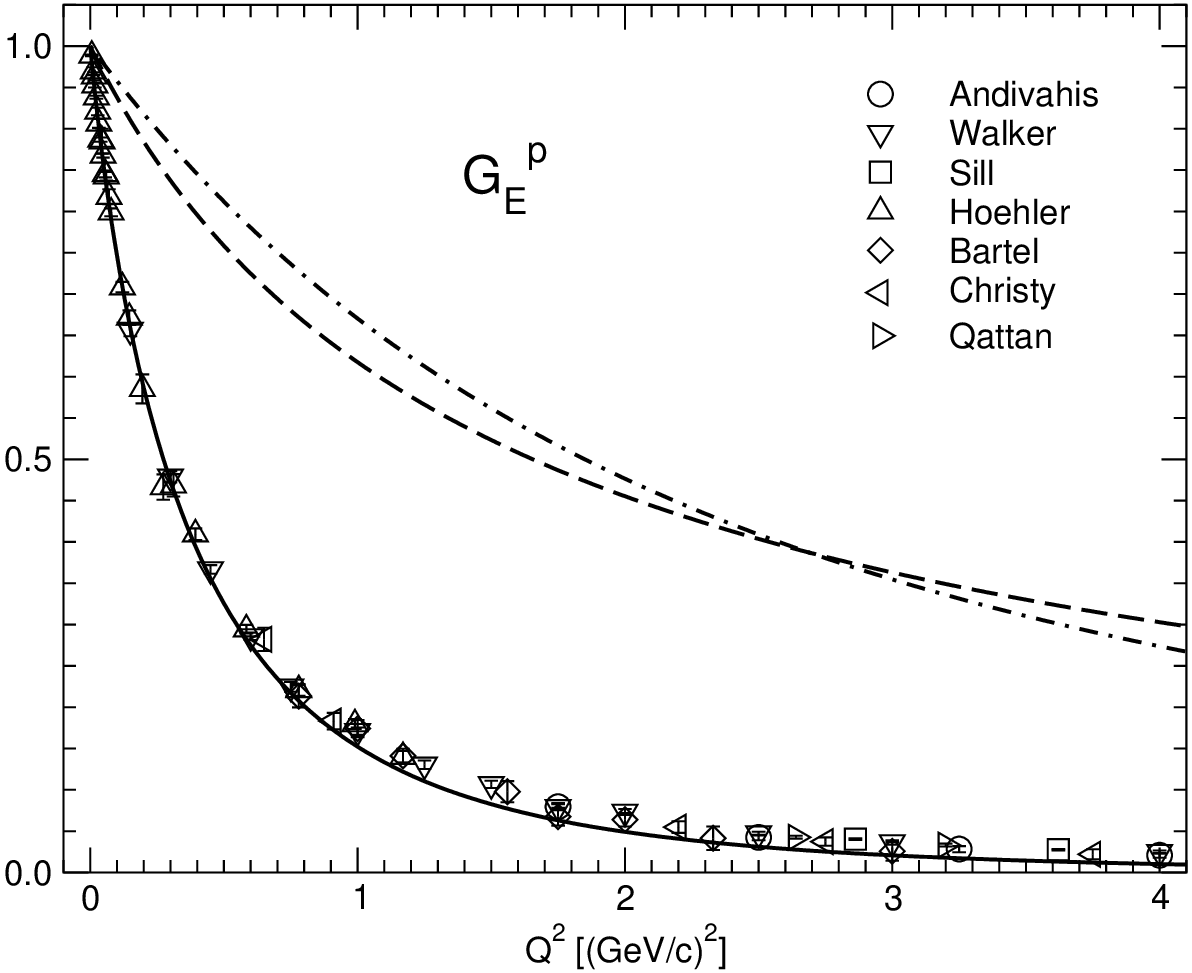}
\hspace{0.3cm}
\includegraphics[clip=,height=6.cm]{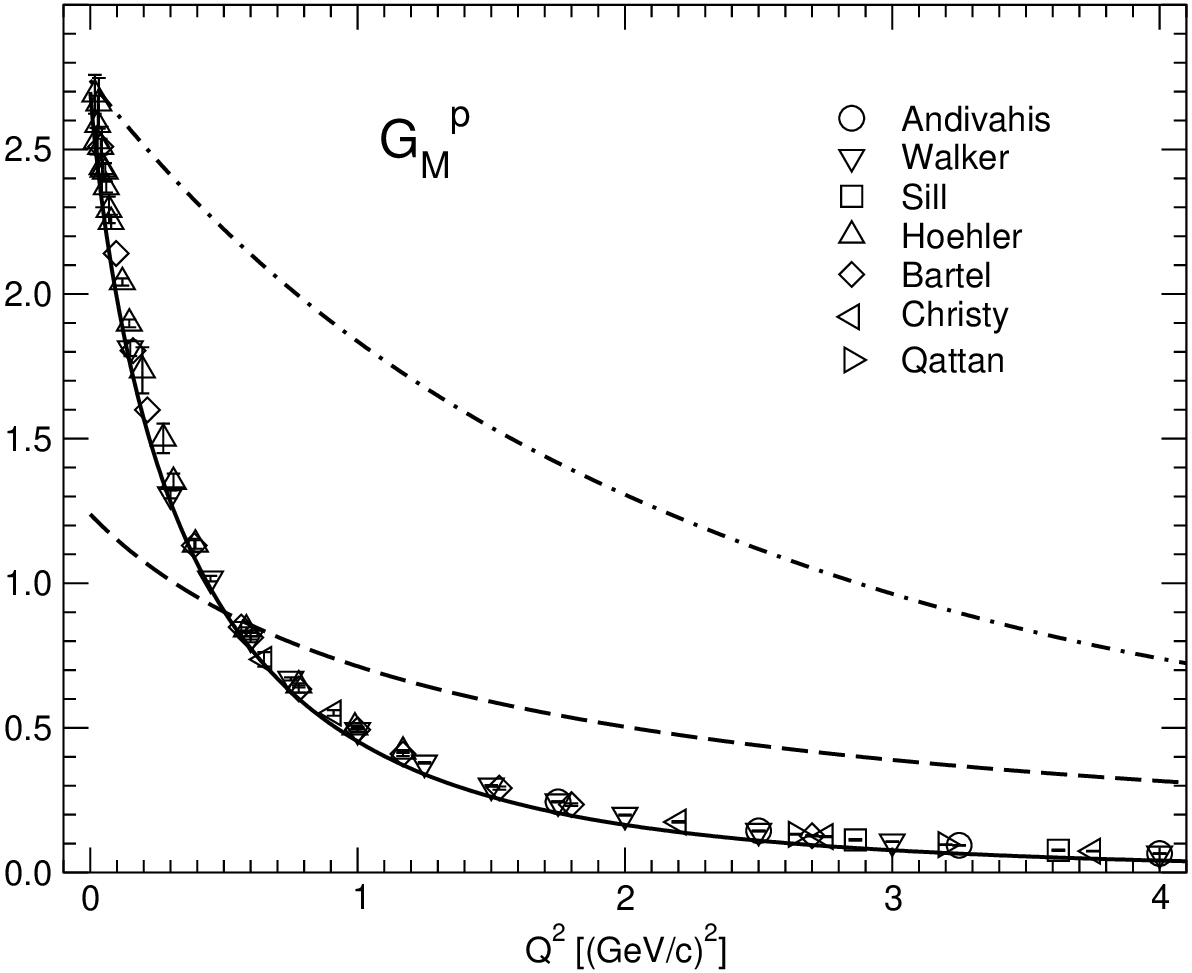}
\caption{\label{fig:proton_sm}
Electric and magnetic form factors of the proton as predicted by the GBE CQM
with the the IFSM (dashed line), PFSM (full line), and the NRIA (dash-dotted line) 
current operators. Experimental data are from
Refs.~\protect\cite{Lung:1993bu,Markowitz:1993hx,%
Rock:1982gf,Bruins:1995ns,Gao:1994ud,Anklin:1994ae,Anklin:1998ae%
,Xu:2000xw,Kubon:2001rj,Xu:2002xc} 
and~\protect\cite{Andivahis:1994rq,Walker:1989af,Sill:1993qw,Hohler:1976ax,Bartel:1973rf,%
Christy:2004rc,Qattan:2004ht}.
}
\end{center}
\end{figure*}
\begin{figure*}[t]
\begin{center}
\includegraphics[clip=,height=6.cm]{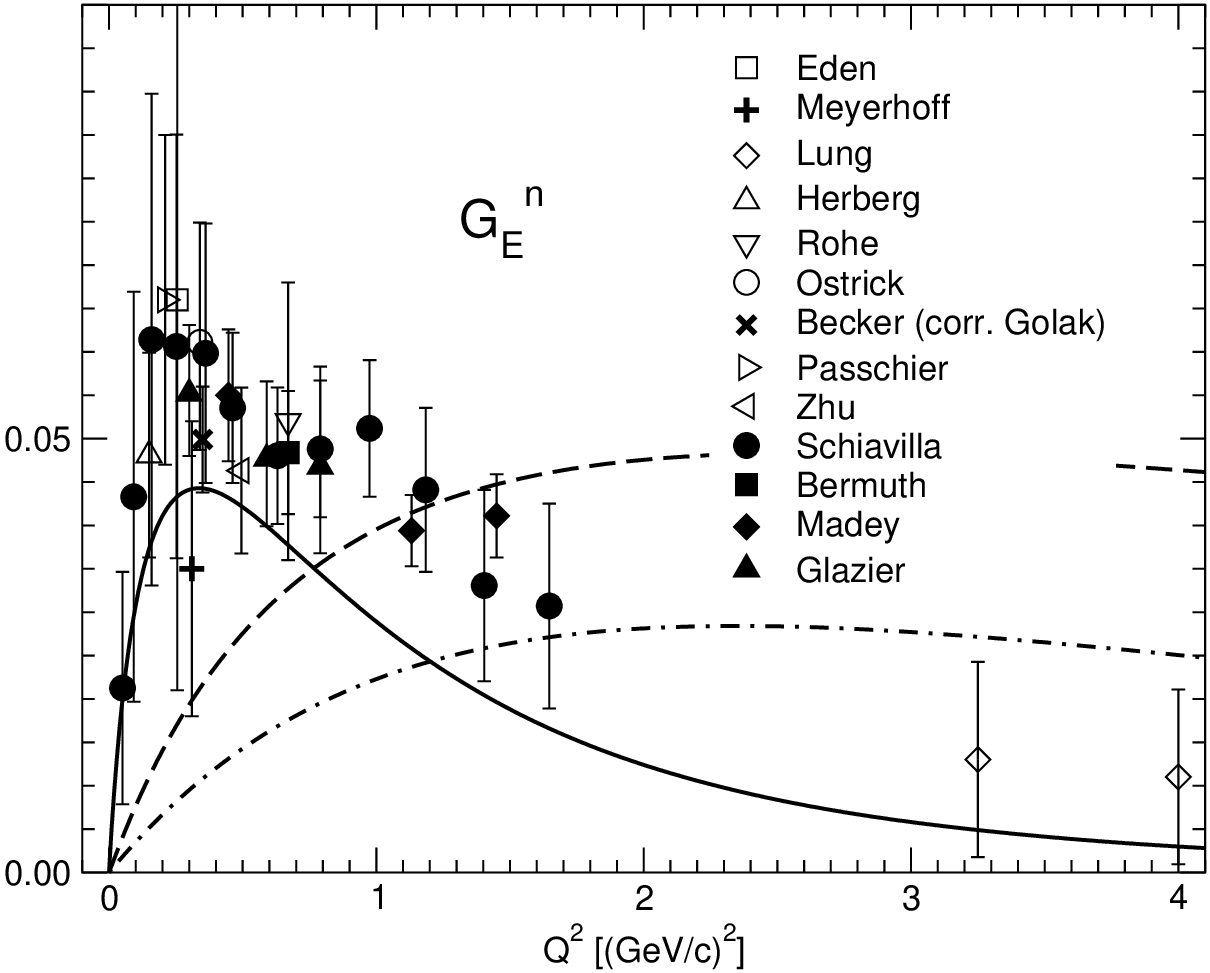}
\hspace{0.3cm}
\includegraphics[clip=,height=6.cm]{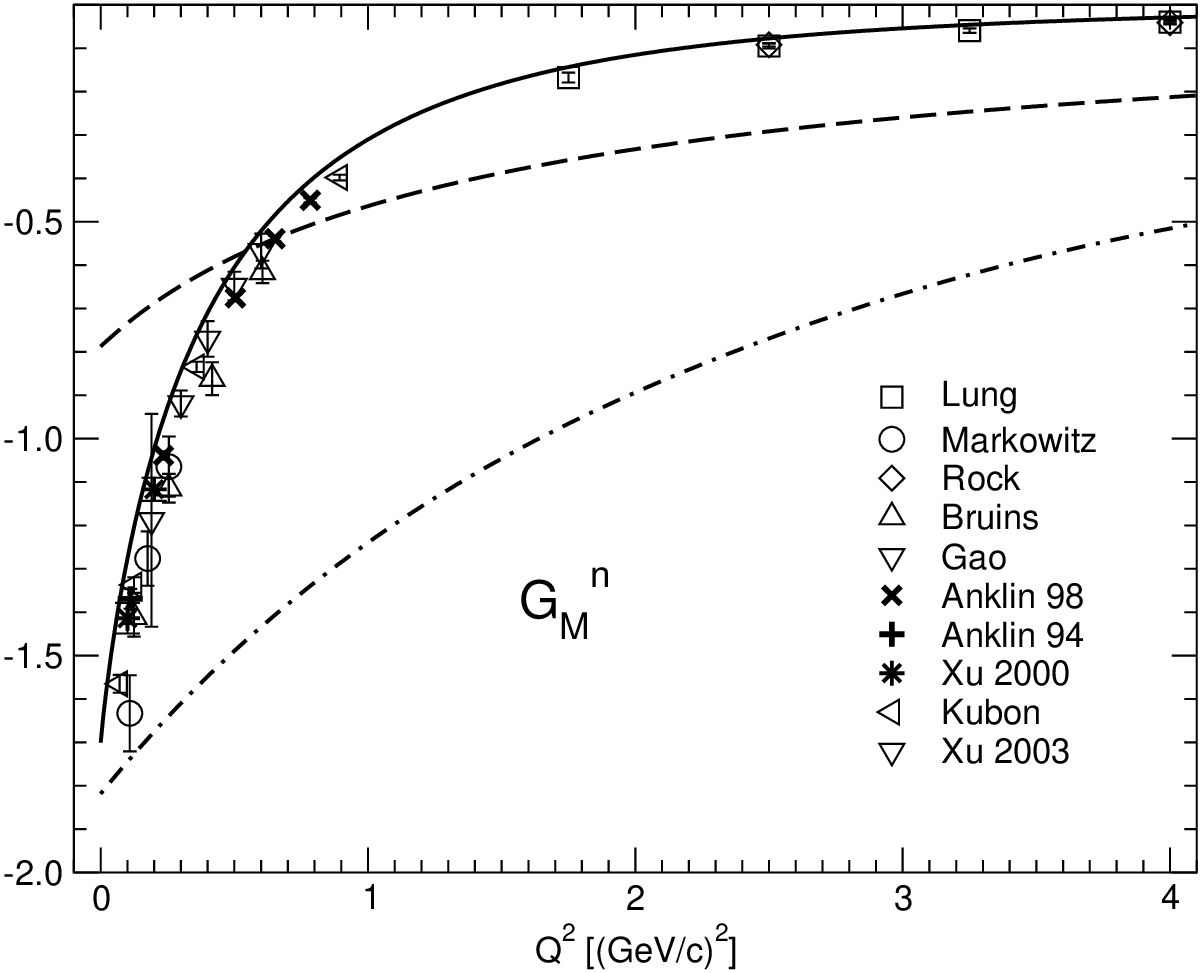}
\caption{\label{fig:neutron_sm}
Same as in Fig.~\ref{fig:proton_sm} but for the neutron.  Experimental data are from
Refs.~\protect\cite{Lung:1993bu,Markowitz:1993hx,%
Rock:1982gf,Bruins:1995ns,Gao:1994ud,Anklin:1994ae,Anklin:1998ae%
,Xu:2000xw,Kubon:2001rj,Xu:2002xc,%
Eden:1994ji,Meyerhoff:1994ev,Herberg:1999ud,Rohe:1999sh,%
Ostrick:1999xa,Becker:1999tw,Passchier:1999cj,Zhu:2001md,%
Schiavilla:2001qe,Bermuth:2003qh,Madey:2003av,Glazier:2004ny}.
}
\end{center}
\end{figure*}
\section{Elastic Nucleon Form Factors}
In this section we present the theoretical results for the elastic electromagnetic
form factors of the nucleon calculated with the spectator-model currents constructed
above. First we compare the IFSM and PFSM results along the GBE CQM. Then we show the
influences from different quark-model wave functions by comparing the GBE CQM results 
with the predictions of the BCN-OGE CQM. We stress that these results have been obtained 
without any parameter variation. However, there reside ambiguities in spectator-model
constructions, which are discussed here quantitatively in case of the PFSM.
\subsection{IFSM versus PFSM results}
Figs.~\ref{fig:proton_sm} and~\ref{fig:neutron_sm} contain the direct predictions
for the Sachs form factors obtained from the GBE CQM. Tables~\ref{tab:magmom}
and~\ref{tab:chargeradii} give the corresponding magnetic moments and charge radii.
Immediately some striking features are evident. For the
electric form factors of both proton and neutron the IFSM results are
very similar to the NRIA and especially in case of the proton, they lie far off the
experimental data. The PFSM predictions fall close to experiment~\cite{Boffi:2001zb}.
For the magnetic form factors the IFSM and NRIA results become quite distinct, especially 
for lower momentum transfers. This has the consequence that the IFSM magnetic moments for
both proton and neutron turn out to be unreasonable in comparison
with experiment, while the NRIA results happen to reproduce them quite well
(see Table~\ref{tab:magmom}). 

Evidently, the spectator-model approximations are different in the instant
and point forms even though they have the same nonrelativistic limit.
The relativistic effects stem from the Lorentz boosts. In PFSM Lorentz boosts
belong to the kinematic subgroup while in IFSM they do not. Consequently
the two forms introduce different effective many-body contributions in the
spectator operators. Therefore,  we may interpret the differences between
the IFSM and PFSM results as being due to different contributions from
(effective) many-body contributions.

\renewcommand{\arraystretch}{1.3}
\begin{table}
\caption{\label{tab:magmom}Magnetic moments of the proton and neutron
(in n.m.) as predicted by the GBE CQM with the the IFSM, PFSM, and the NRIA
current operators. Experimental data after the PDG~\cite{PDBook}.}
\begin{ruledtabular}
\begin{tabular}{crrrr}
&\multicolumn{3}{c}{GBE CQM}&Experiment\\
Nucleon & IFSM & PFSM &  NRIA & \\
\hline
 p & 1.24 & 2.70 & 2.74 & 2.79 \\
 n & -0.79 & -1.70 &-1.82 & -1.91
 \end{tabular}
\end{ruledtabular}
\end{table}
\begin{table}
\caption{\label{tab:chargeradii}Charge radii of the proton and neutron
(in fm$^2$) as predicted by the GBE CQM with the the IFSM, PFSM, and the NRIA
current operators. Experimental data after the PDG~\cite{PDBook}.}
\begin{ruledtabular}
\begin{tabular}{crrrr}
&\multicolumn{3}{c}{GBE CQM}&Experiment\\
Nucleon & IFSM & PFSM & NRIA & \\
\hline
 p& 0.156 & 0.824  & 0.102 & 0.766  \\
 n & -0.020 &  -0.135 & -0.009 & -0.116
 \end{tabular}
\end{ruledtabular}
\end{table}

It should be emphasized that we make the comparison of the IFSM, PFSM, and NRIA
predictions without any readjustment of the CQM parameters. Here,
we directly employ the nucleon wave function as produced by the GBE CQM, whose
parameters were fitted only to the baryon spectra in Ref.~\cite{Glozman:1998ag}.
One could bring, for example, the IFSM results closer to experiment by an ad-hoc
modification of the nucleon wave function and an adjustment of the constituent 
quark mass, as is done in Ref.~\cite{Julia-Diaz:2003gq}. 
However, this would then change also the PFSM and NRIA results considerably,
making a consistent comparison of the CQM predictions obtained with the
various spectator-model constructions difficult if not impossible.
\subsection{Effects from Quark-Model Wave Functions}
\label{sec:5}
Having compared the results with different spectator-model constructions of the
EM current (in case of the GBE CQM), we are now interested to see
the effects from different relativistic CQM nucleon wave functions. This comparison 
is performed along the PFSM approach. In addition
to the GBE CQM (whose hyperfine interaction is flavour dependent) we have 
calculated the predictions of a different relativistic CQM with a chromomagnetic 
hyperfine interaction based on OGE dynamics. 
In particular, we have employed the relativistic variant of the
BCN CQM~\cite{Bhaduri:1981pn} in the parametrization of
Ref.~\cite{Theussl:2000sj}. For completeness we have also included the
predictions by even another type of relativistic CQM, namely the II CQM by
the Bonn group~\cite{Loring:2001kx,Loring:2001ky}. These results, however,
stem from the field-theoretic Bethe-Salpeter approach which is principally
different from the RQM we followed. The current operator
employed in Ref.~\cite{Merten:2002nz} has also been approximated by a
spectator-model construction.

\begin{figure*}[t]
\begin{center}
\includegraphics[height=6.0cm,clip=]{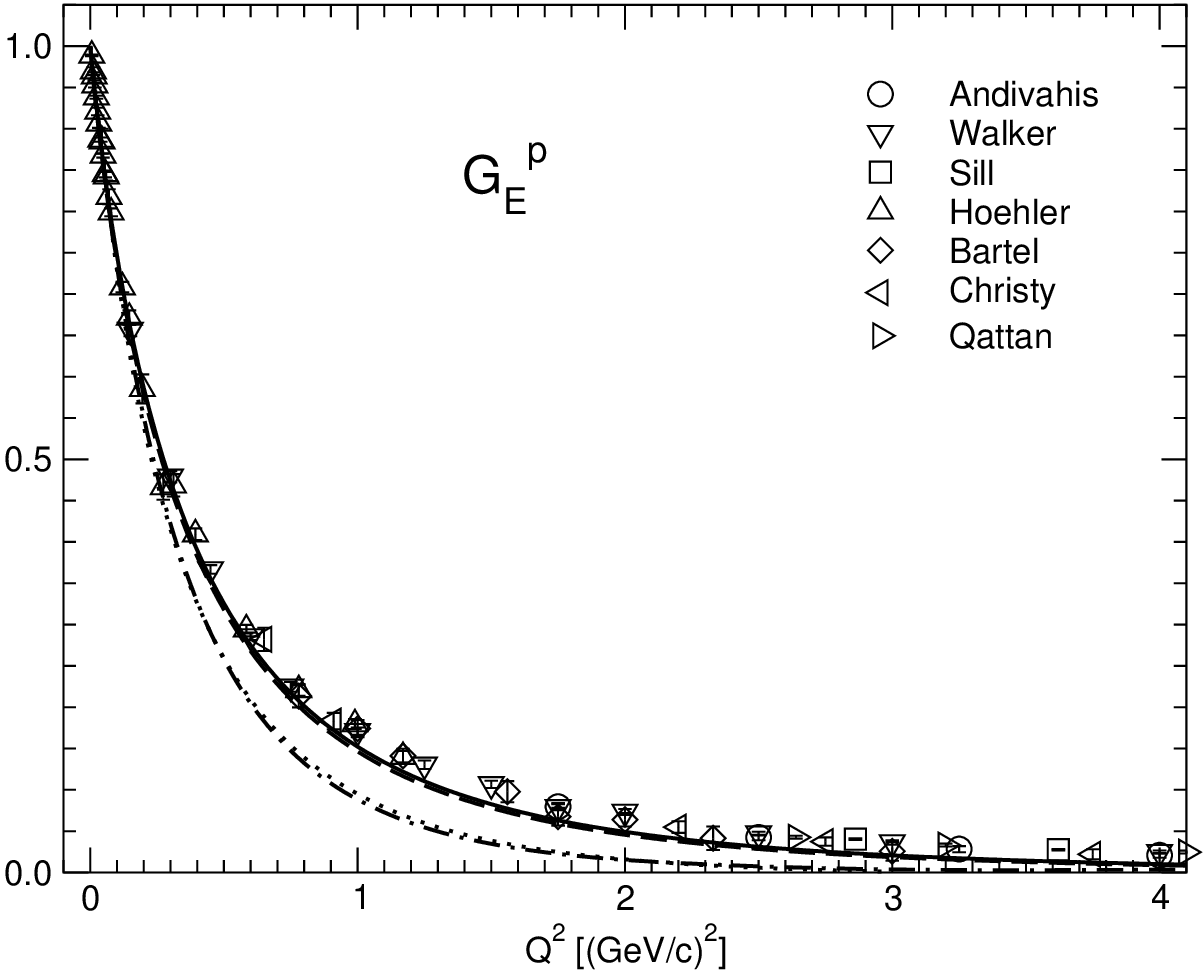}
\hspace{0.3cm}
\includegraphics[height=6.0cm,clip=]{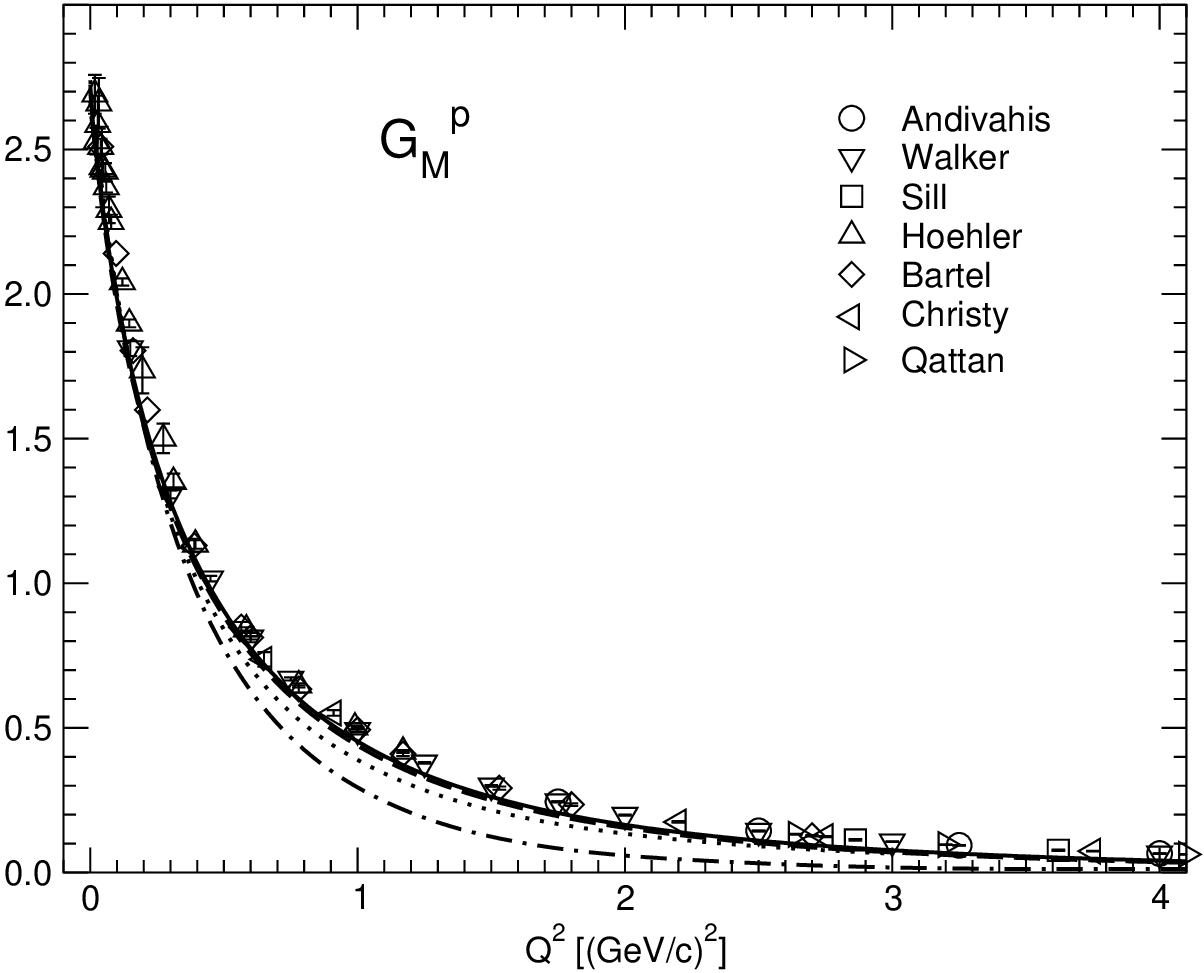}
\caption{\label{fig:proton_comp}
Electric and magnetic form factors of the proton as predicted by the
GBE (full line) and BCN (dashed line) CQMs along the PFSM approach; in addition
the results for the case with only the confinement potential (inherent in the
GBE CQM) are given (dash-dotted line). For comparison also the predictions of
the II CQM (dotted line) after Ref.~\cite{Merten:2002nz} are shown.
Experimental data as in Fig.~\ref{fig:proton_sm}.
}
\end{center}
\end{figure*}
\begin{figure*}[t]
\begin{center}
\includegraphics[height=6.0cm,clip=]{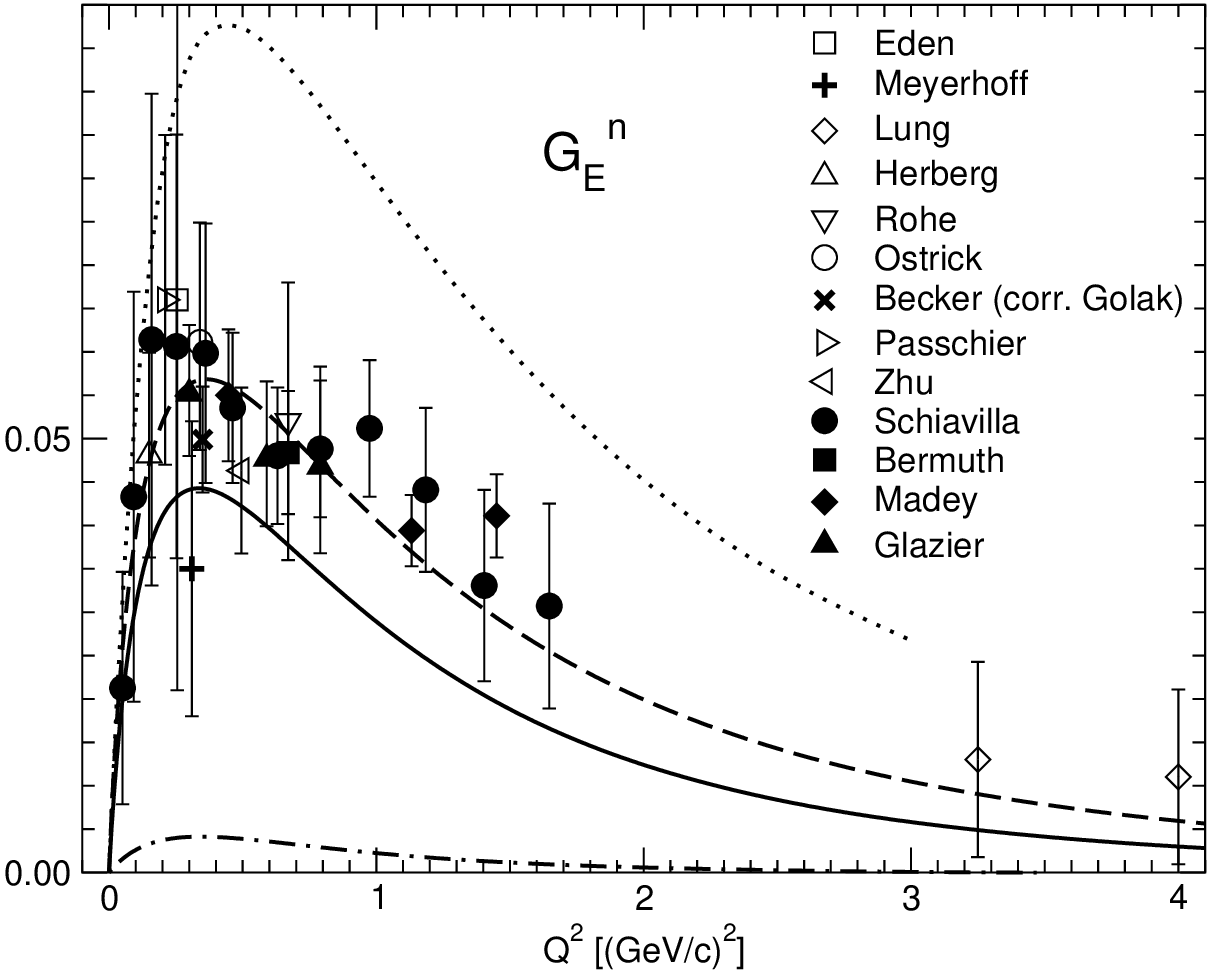}
\hspace{0.3cm}
\includegraphics[height=6.0cm,clip=]{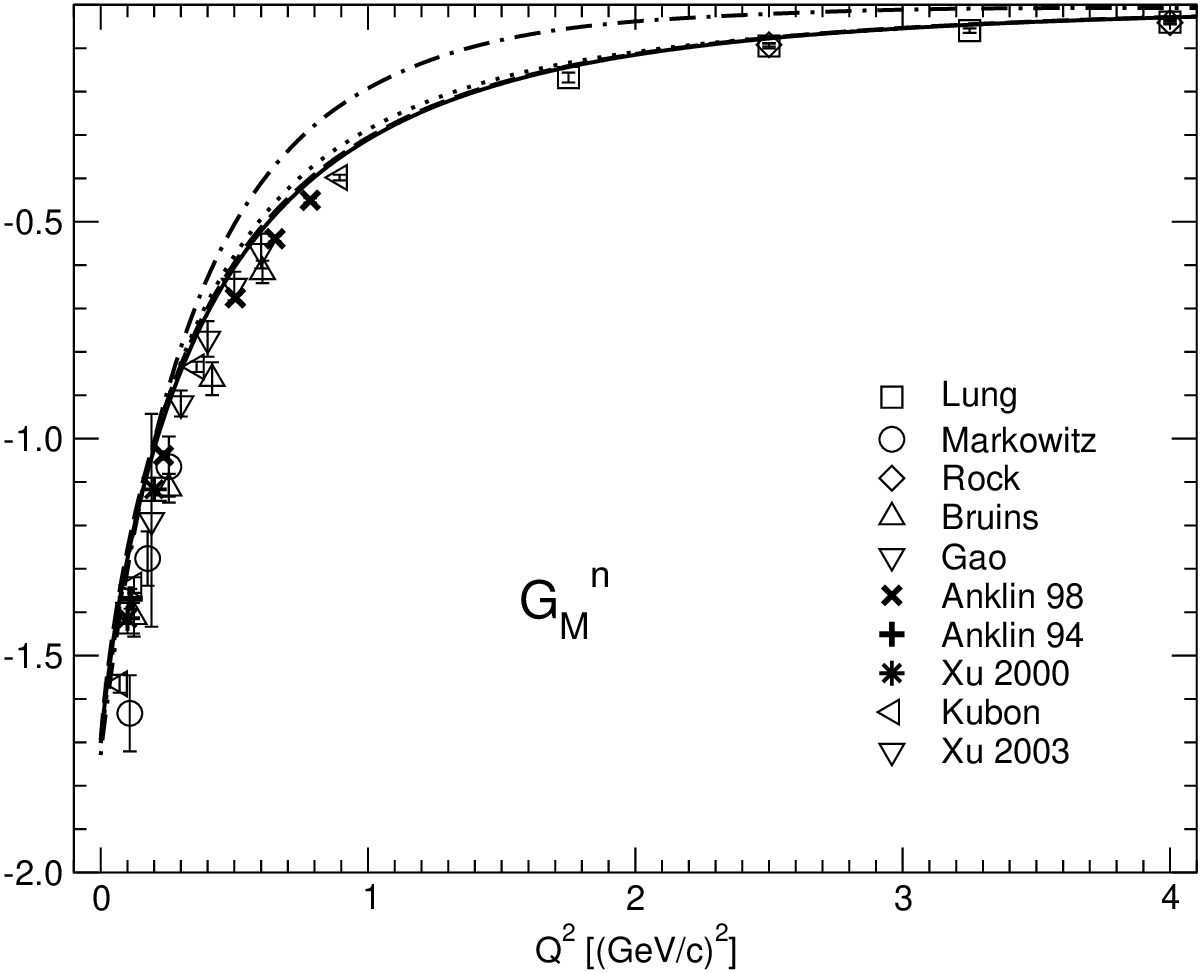}
\caption{\label{fig:neutron_comp}
Same as in Fig.~\ref{fig:proton_comp} but for the neutron. Experimental data as in
Fig.~\ref{fig:neutron_sm}.
}
\end{center}
\end{figure*}

In Figs~\ref{fig:proton_comp} and~\ref{fig:neutron_comp} we compare the 
predictions for the EM form factors of proton and neutron, respectively; 
the corresponding results for the electric
radii and magnetic moments are quoted in Tables~\ref{tab:magmom_comp}
and~\ref{tab:chargeradii_comp}. 
One observes that the differences between the GBE and BCN CQMs are relatively small;
in most cases the curves are even indistinguishable.
Only for the electric form factor of the neutron the discrepancies between the curves
are clearly visible, but here one should take into account the expanded scale in 
that figure. Differences between the GBE and BCN CQMs are also
found in the charge radii, whereas the magnetic moments are again very similar.
In all instances, the PFSM predictions of the GBE and OGE CQMs are, nevertheless,
found in remarkable vicinity of the experimental data, except for the electric radius
of the neutron in case of the BCN CQM. Certainly, the
differences between the two types of CQMs are much smaller than the variations
found above between the IFSM and PFSM results. 

It is now interesting to compare also with the predictions of the II CQM.
Surprisingly, they are very similar to the ones of the GBE and OGE CQMs, even
though they have been derived in a completely different framework, namely
along the Bethe-Salpeter approach. Only, the results
from the II CQM tend to undershoot the proton electric form factor and
overestimate the neutron electric form factor but generally there is a similarity
of the Bethe-Salpeter and PFSM results.
  
On the other hand, a CQM, where only the confinement potential is present, fails in 
reproducing the electromagnetic nucleon structure. In particular, it yields the
proton form factors too small and misses the neutron electric
form factor completely. The reason is that a certain mixed-symmetric 
spatial component in the neutron wave function is needed in order to produce
a non-vanishing electric form factor. However, the neutron wave function produced
by the confinement-only potential comes practically without a mixed-symmetric spatial
part. That such type of nucleon wave function cannot be adequate is consistent with
observations made already in earlier works~\cite{sprung,chernyak}.
\begin{table}
\caption{\label{tab:magmom_comp}Magnetic moments of the proton and neutron
(in n.m.) as predicted by the GBE, BCN and confinement-only (CONF) CQMs along 
the PFSM and the II CQM along the Bethe-Salpeter (BS) approach. 
Experimental data after the PDG~\cite{PDBook}.}
\begin{ruledtabular}
\begin{tabular}{crrrrr}
&\multicolumn{3}{c}{PFSM}&BS&Experiment\\
Nucleon & GBE & BCN &  CONF & II & \\
\hline
 p & 2.70 & 2.74 & 2.65 & 2.74 & 2.79 \\
 n & -1.70 & -1.70 & -1.73 & -1.70 & -1.91
 \end{tabular}
\end{ruledtabular}
\end{table}
\begin{table}
\caption{\label{tab:chargeradii_comp}Charge radii of the proton and neutron
(in fm$^2$) as predicted by the GBE, BCN and confinement-only (CONF) CQMs along 
the PFSM and the II CQM along the Bethe-Salpeter (BS) approach. 
Experimental data after the PDG~\cite{PDBook}.}
\begin{ruledtabular}
\begin{tabular}{crrrrr}
&\multicolumn{3}{c}{PFSM}&BS&Experiment\\
Nucleon & GBE & BCN &  CONF & II & \\
\hline
 p& 0.824 & 1.029  & 0.766 & 0.67 & 0.766  \\
 n & -0.135 &  -0.263 & -0.009 & -0.11 & -0.116
 \end{tabular}
\end{ruledtabular}
\end{table}
\begin{figure*}[t]
\begin{center}
\includegraphics[height=6.0cm,clip=]{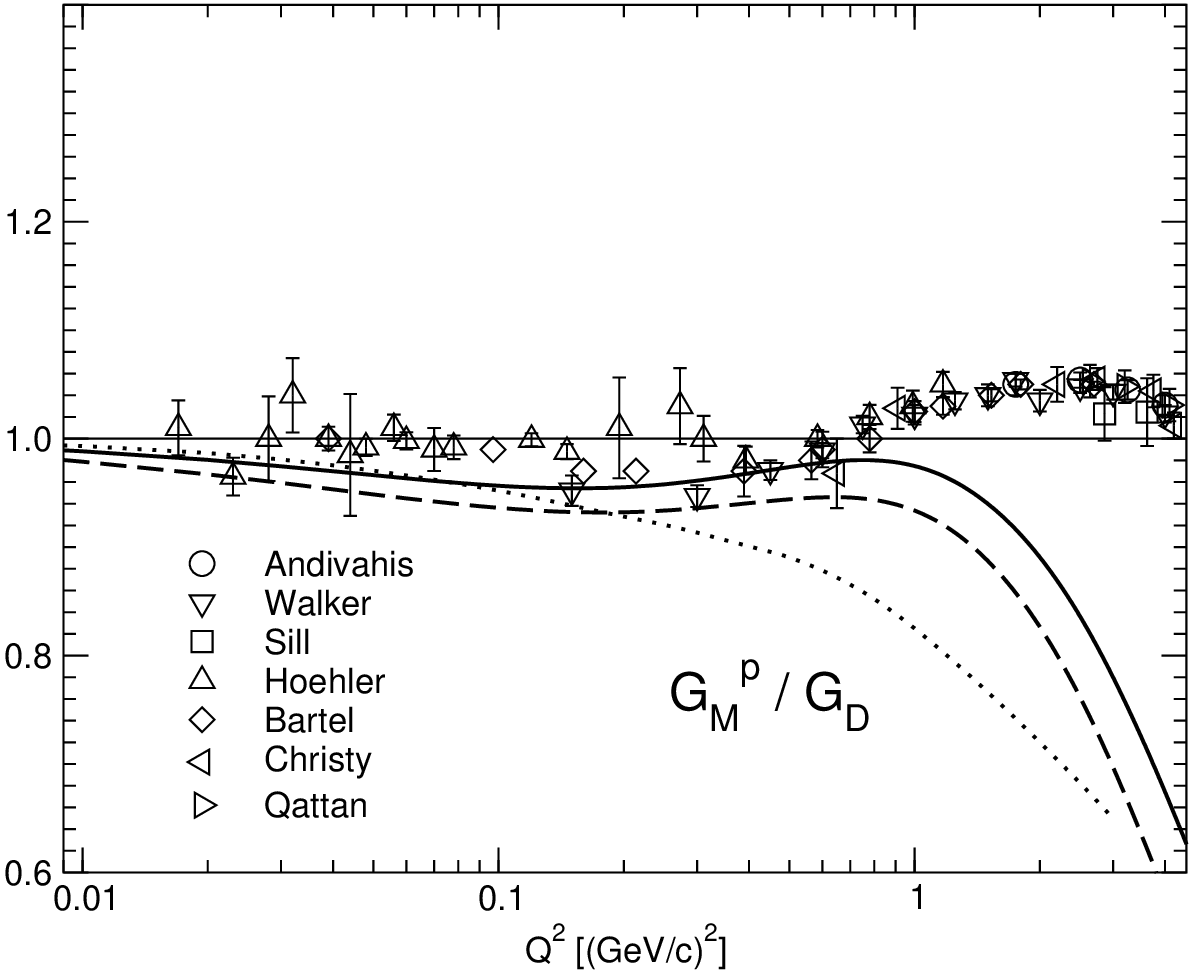}
\hspace{0.3cm}
\includegraphics[height=6.0cm,clip=]{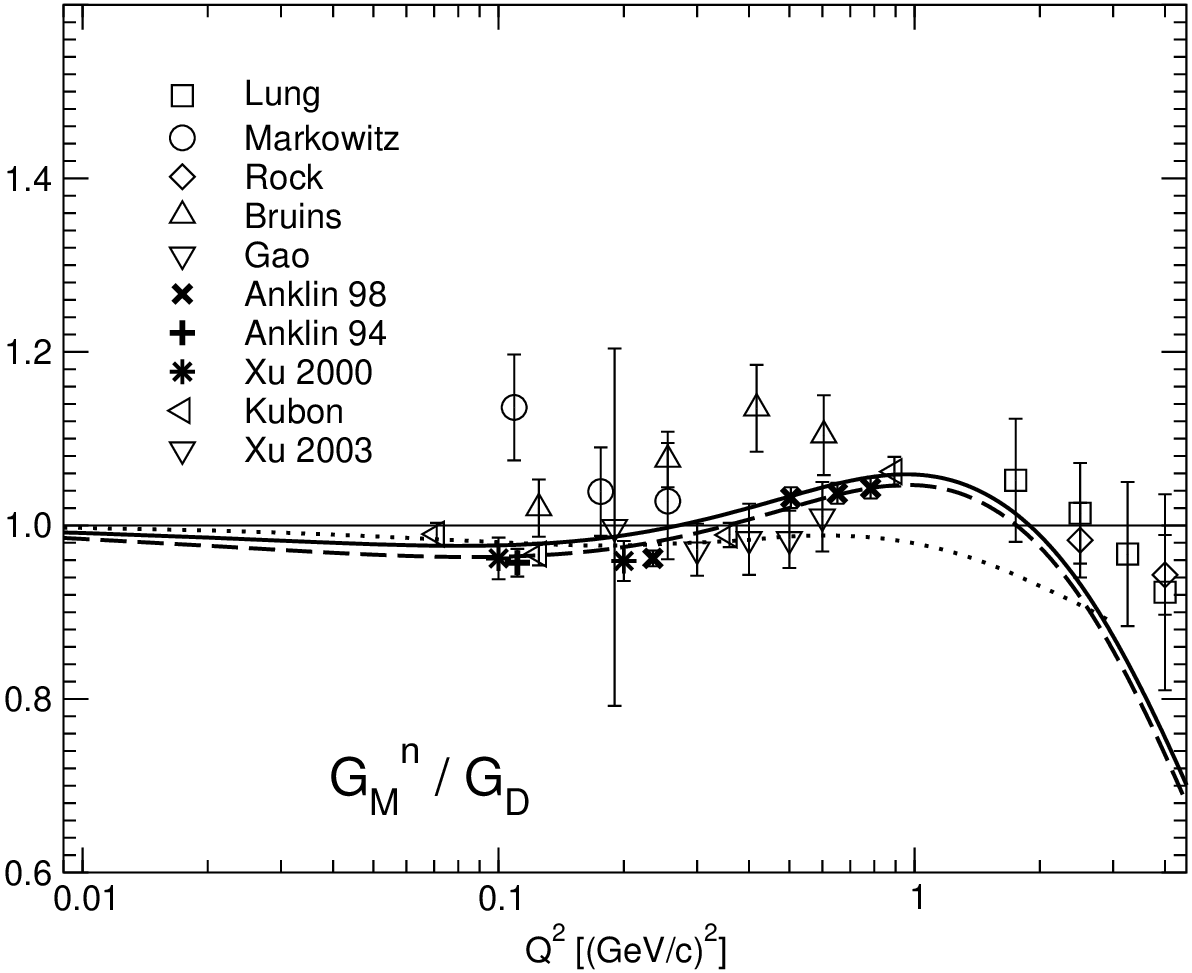}
\caption{\label{fig:CQM ratio}
Same comparison as in Figs.~\ref{fig:proton_comp} and~\ref{fig:neutron_comp}
for the ratios of magnetic to dipole form factors of the proton (left panel)
and the neutron (right panel). All ratios are normalized to $1$ at $Q^2=0$.
}
\end{center}
\end{figure*}

For a closer inspection of the differences in the relativistic CQM predictions
it is instructive to look
through the magnifying glass of $G_M^n/G_D$ and $G_M^p/G_D$ ratios. This 
comparison is given in Fig.~\ref{fig:CQM ratio}. The tiny differences between the PFSM
predictions of the GBE and OGE CQMs are confirmed. The $Q^2$ dependence of the 
ratios is seen to be slightly distinct in the case of the II CQM.

From the comparisons made in this subsection we conclude that realistic nucleon
wave functions are necessary in order to describe the proton as well as neutron 
electromagnetic structure consistently. Such type of wave functions are
obviously achieved in the GBE, OGE, and II CQMs. With respect to the nucleon
electromagnetic structure the presence of a hyperfine interaction itself is found of considerable
importance. Even though the various hyperfine interactions of the CQMs considered
here stem from different dynamics, they obviously lead to similar nucleon wave
functions. When the hyperfine
interaction is left out completely (cf. the case with the confinement potential only),
the nucleon structure can by no means be described in a reasonable manner.

\subsection{Uncertainties in the PFSM Construction}
\label{sec:3}
Finally we deal with the problem that the PFSM construction 
has some residual ambiguity. In particular, a factor ${\cal N}$ necessarily appears in
the PFSM current of Eq.~(\ref{eq:pfsm}). As already mentioned the factor ${\cal N}$ is
unavoidable, if one wants the spectator model current to reduce to a genuine
one-body operator in the limit of zero-momentum
transfer~\cite{Melde:2004qu}. It is also required to guarantee for the proper
charge normalization of the proton.

All results for the electromagnetic nucleon structure considered so far
have been calculated with the symmetric choice for the factor ${\cal N}$
as given in Eq.~(\ref{eq:geoN}). However, this is not the only possibility.
Following Ref.~\cite{Melde:2004qu} one could also adopt the
general expression
\begin{equation}
{\cal N}\left(x,y\right)=
\left(\frac{M}{\sum_i{\omega_{i}}}\right)^{xy}
\left(\frac{M}{\sum_i{\omega'_{i}}}\right)^{x\left(1-y\right)}\, ,
\label{eq:offsymfac}
\end{equation}
where $x$ and $y$ are to be considered as open parameters varying in the 
range $0\le x$ and $0\le y\le 1$. The normalization
factors so defined are all Lorentz invariant and all lead to a covariant 
PFSM current with the required properties.

\begin{figure}[t]
\begin{center}
\includegraphics[clip=,width=7.2cm]{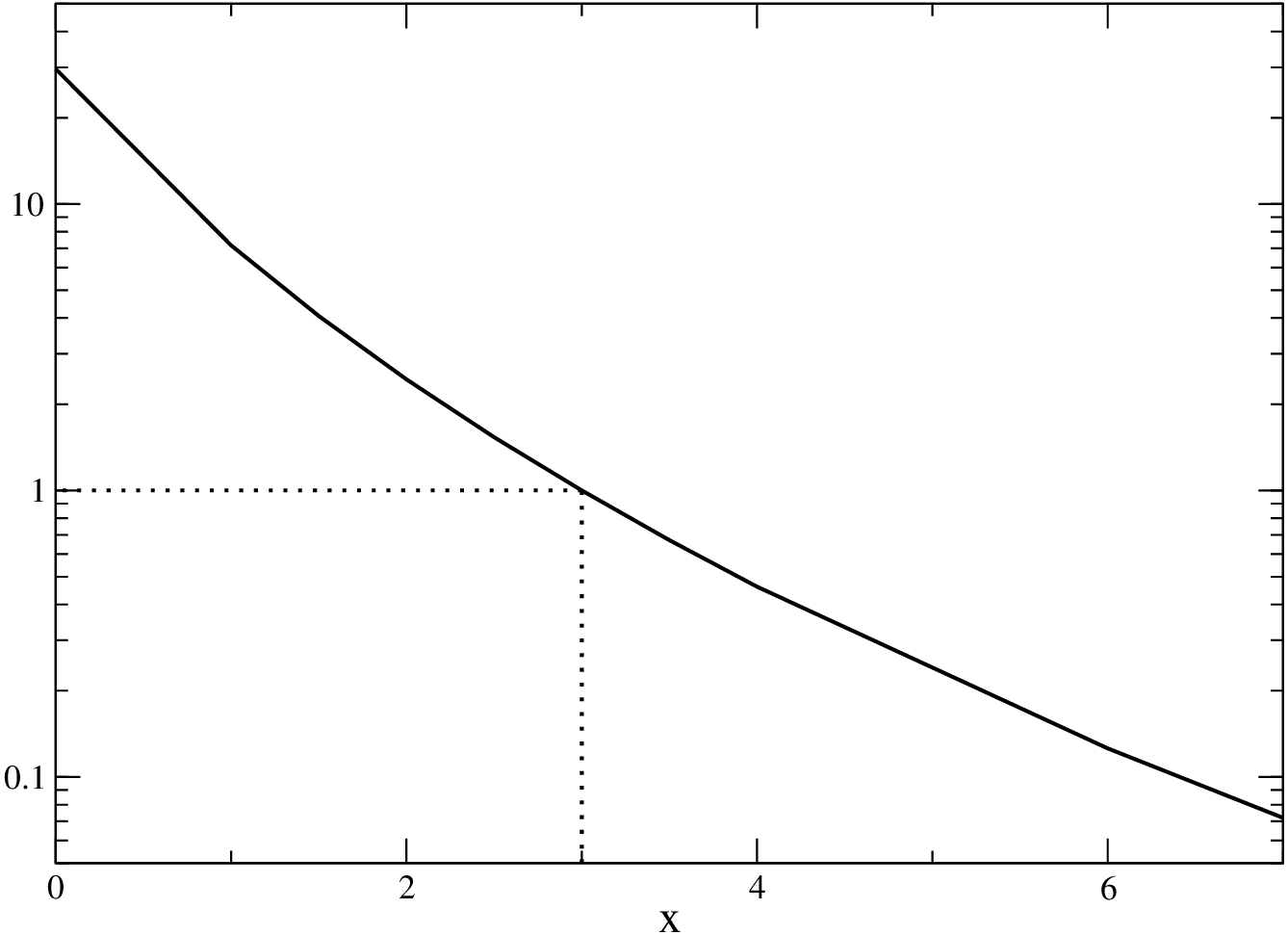}
\caption{\label{fig:chargenorm}
Electric proton form factor $G_E^p$ at momentum transfer $Q^2=0$ (i.e. proton charge) 
as a function of the exponent parameter $x$ in the PFSM factor
$\cal N$ of Eq.~(\ref{eq:offsymfac}), calculated with the GBE CQM~\cite{Glozman:1998ag}.
}
\end{center}
\end{figure}

In the following, we examine the dependence of the EM form factor results on
possible choices of $\cal N$, i.e. variations of the parameters $x$ and $y$.
At zero momentum transfer the factor $\cal N$ depends only on $x$. 
In Fig.~\ref{fig:chargenorm} we show the results for the electric proton
form factor $G_{\rm E}^p\left(Q^2\right)$ at $Q^2=0$, i.e. the
proton charge, for varying $x$. Clearly, the only possibility consistent with
the proton charge normalization is $x=3$. 
This result is in line also with the arguments given in Ref.~\cite{Melde:2004qu} 
that in the limit $Q^2\rightarrow 0$ the PFSM reduces to a genuine one-body
operator only with a cubic choice for the factor ${\cal N}$.

\begin{figure}[t]
\begin{center}
\includegraphics[clip=,width=8.1cm]{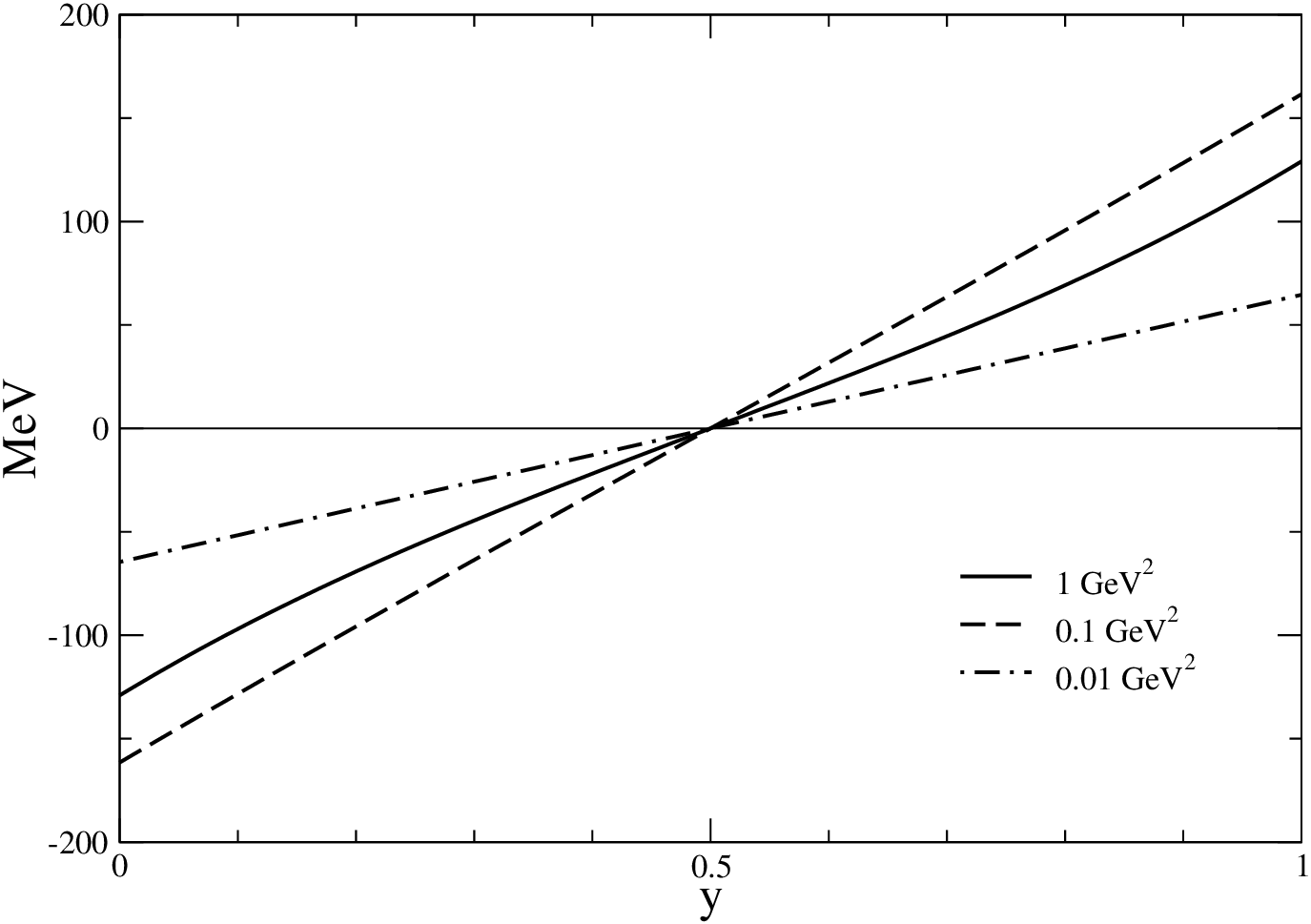}
\caption{\label{fig:currentcons}
Third component of the transition amplitude
$F^{3}_{\frac{1}{2},\frac{1}{2}}\left(Q^2\right)$ in the Breit frame as a function
of the exponent parameter $y$ in the PFSM factor
$\cal N$ of Eq.~(\ref{eq:offsymfac}) for three different values of the momentum
transfer $Q^2$, calculated with the GBE CQM~\cite{Glozman:1998ag}.
}
\end{center}
\end{figure}

Once the parameter $x$ is uniquely fixed, let us now examine the possible $y$-dependence.
In Fig.~\ref{fig:currentcons} we show 
the third component of the transition amplitude~(\ref{eq:FmuPF})
in the Breit frame as a function of $y$ for three different momentum transfers. It is
well known that in the Breit frame the expectation value of the
third component of the current has to vanish
under the constraint of time-reversal invariance~\cite{Durand:1962,Ernst:1960}.
From the results in Fig.~\ref{fig:currentcons} it is immediately seen that there
is a unique value of $y$ where the third component of the transition amplitude
vanishes for all values of $Q^2$. Consequently, $y=\frac{1}{2}$ has to
be chosen in Eq.~(\ref{eq:offsymfac}) in order to satisfy time-reversal invariance.
This motivates the symmetric choice in Eq.~(\ref{eq:geoN}).

However, we may even find another choice for the factor $\cal N$ that meets all
requirements posed, including time-reversal invariance. A valid construction would also be
\begin{widetext}
\begin{equation}
{\cal N}\left(z\right)=
\frac{1}{2}\left[
\left(\frac{M}{\sum_i{\omega_{i}}}\right)^{3z}
\left(\frac{M}{\sum_i{\omega'_{i}}}\right)^{3\left(1-z\right)}
+
\left(\frac{M}{\sum_i{\omega_{i}}}\right)^{3\left(1-z\right)}
\left(\frac{M}{\sum_i{\omega'_{i}}}\right)^{3z}
\right]
\, .
\label{eq:facgen}
\end{equation}
\end{widetext}
with an open parameter $z$ varying in the range $0\le z\le 1$. This expression
produces the symmetric choice of Eq.~(\ref{eq:geoN}) for the special value
$z=\frac{1}{2}$.

Another form is obtained by choosing $z=0$ (or equivalently $z=1$),
leading to
\begin{equation}
{\cal N}_{\rm ari}=
\frac{1}{2}\left[
\left(\frac{M}{\sum_i{\omega_{i}}}\right)^{3}+
\left(\frac{M}{\sum_i{\omega'_{i}}}\right)^{3}
\right]\, ,
\label{eq:arithfac}
\end{equation}
which can be viewed as the arithmetic mean of two pieces relating to the
ratios of the interacting and free masses in the incoming and outgoing nucleon
states.

Evidently, all of the allowed forms of $\cal N$, according to Eqs.~(\ref{eq:offsymfac})
and (\ref{eq:facgen}), lead to the same nonrelativistic limit, which consists in the NRIA.

\begin{figure*}[t]
\begin{center}
\includegraphics[clip=,height=6.cm]{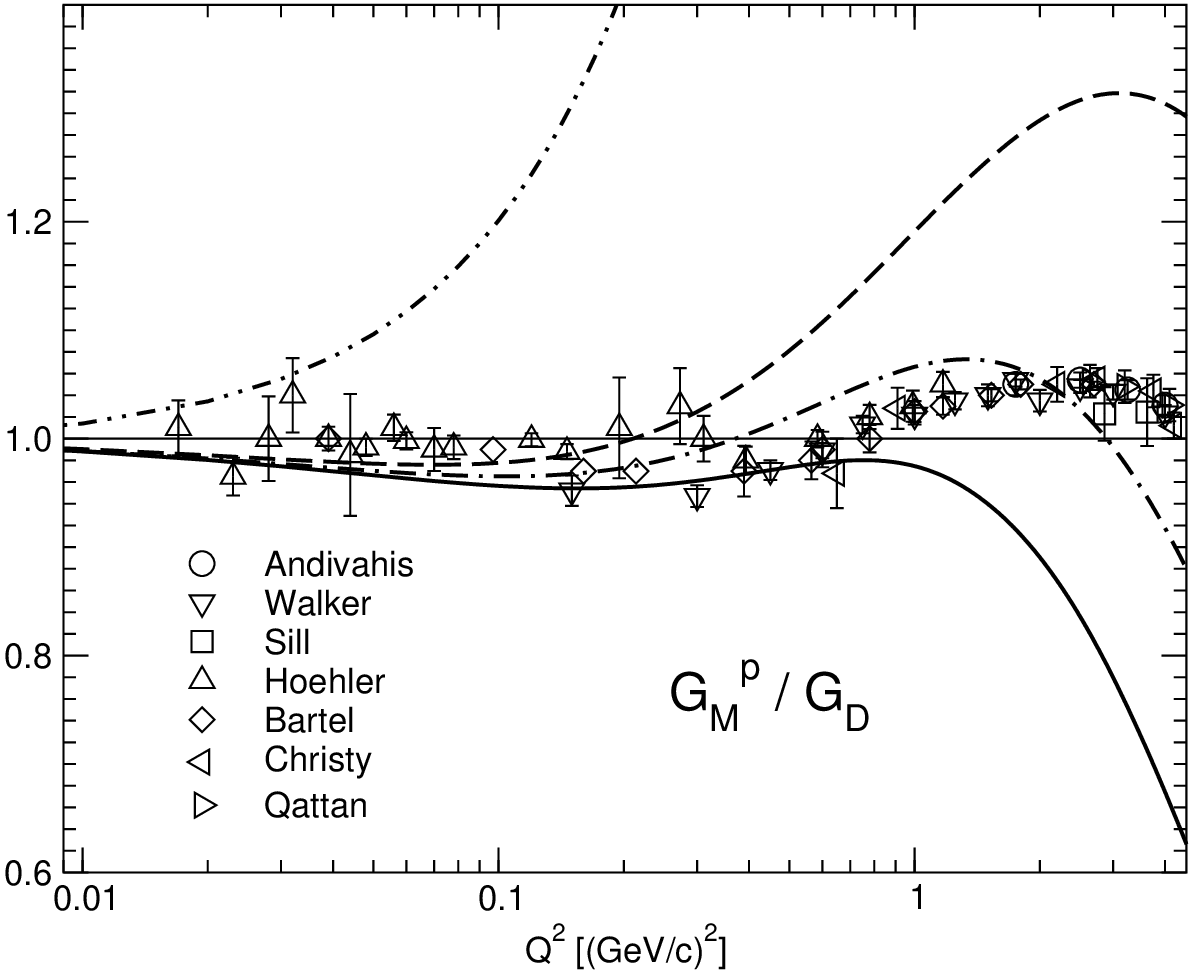}
\hspace{0.3cm}
\includegraphics[clip=,height=6.cm]{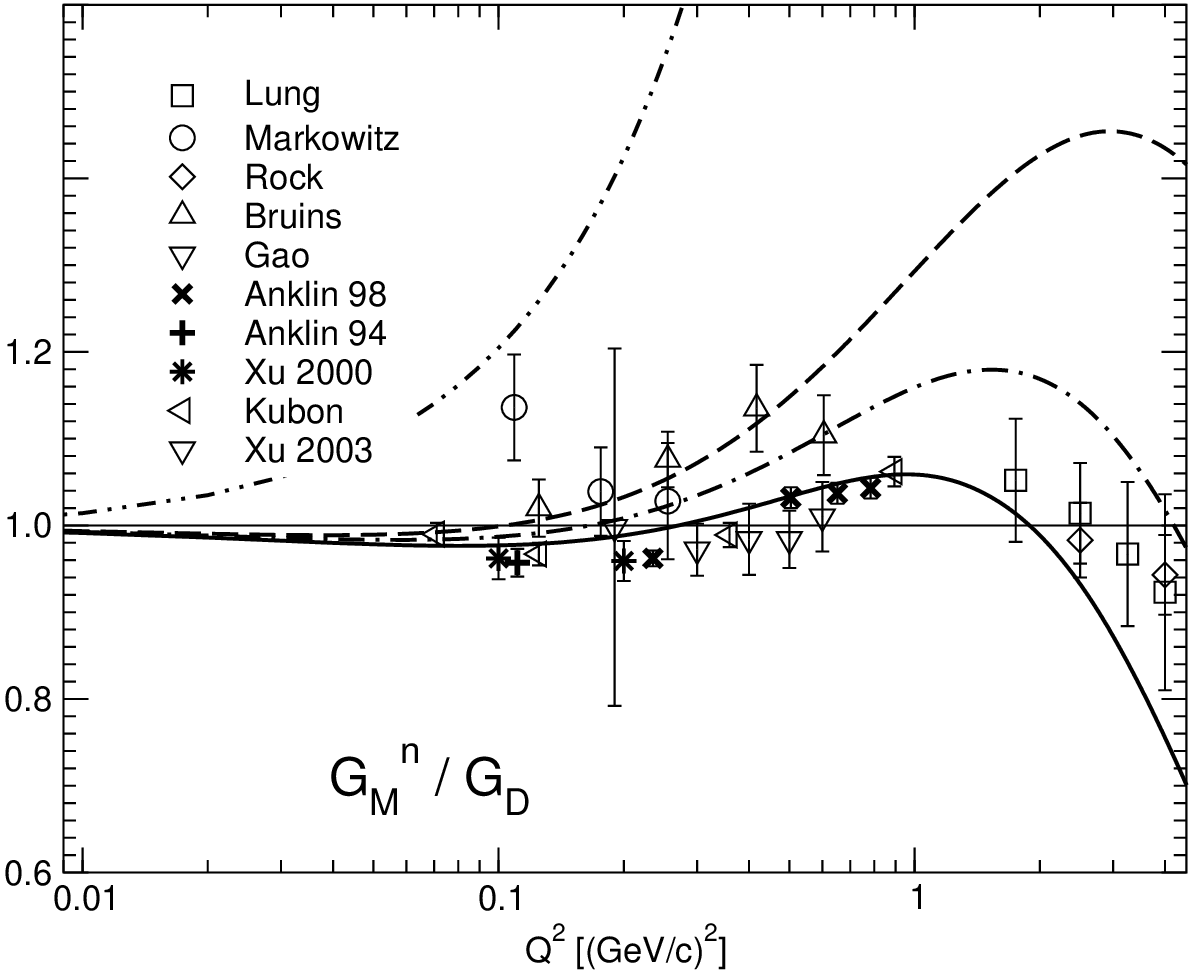}
\caption{
\label{fig:ratio_ari_geo}
Ratios of magnetic form factor to standard dipole paramterization for the 
proton (left) and neutron (right) with different PFSM currents in case of
the GBE CQM.
The full lines denote the results with ${\cal N}_{\rm S}$ from the previous
subsections, the dashed lines with ${\cal N}_{\rm ari}$ from Eq.~(\ref{eq:arithfac}),
and the dash-dotted lines with ${\cal N}_{\rm fit}$, see the text. For
comparison also the IFSM results are shown (dash-double-dotted lines).
All ratios are normalized to $1$ at $Q^2=0$.
}
\end{center}
\end{figure*}

\begin{figure*}[t]
\begin{center}
\includegraphics[clip=,height=6.cm]{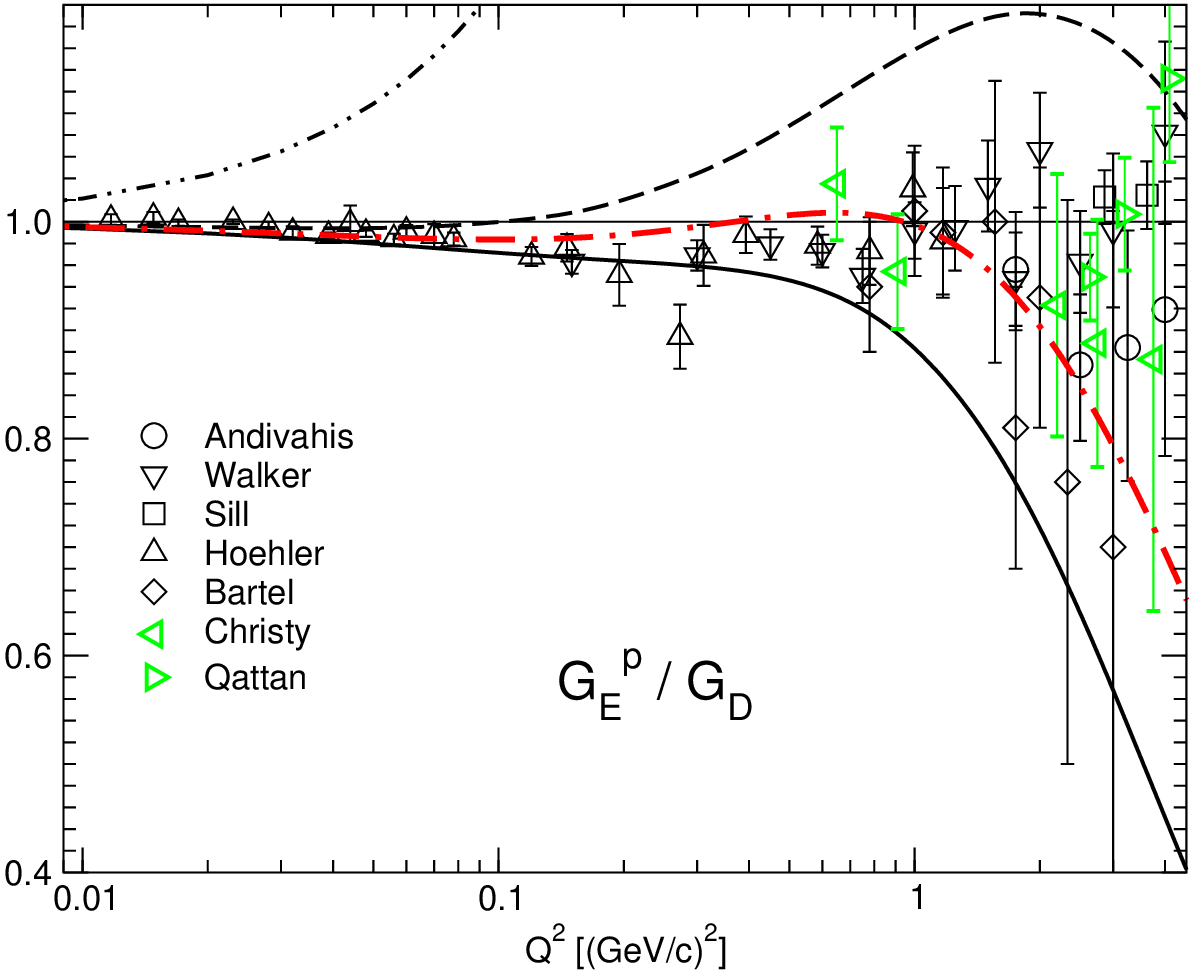}
\hspace{0.3cm}
\includegraphics[clip=,height=6.cm]{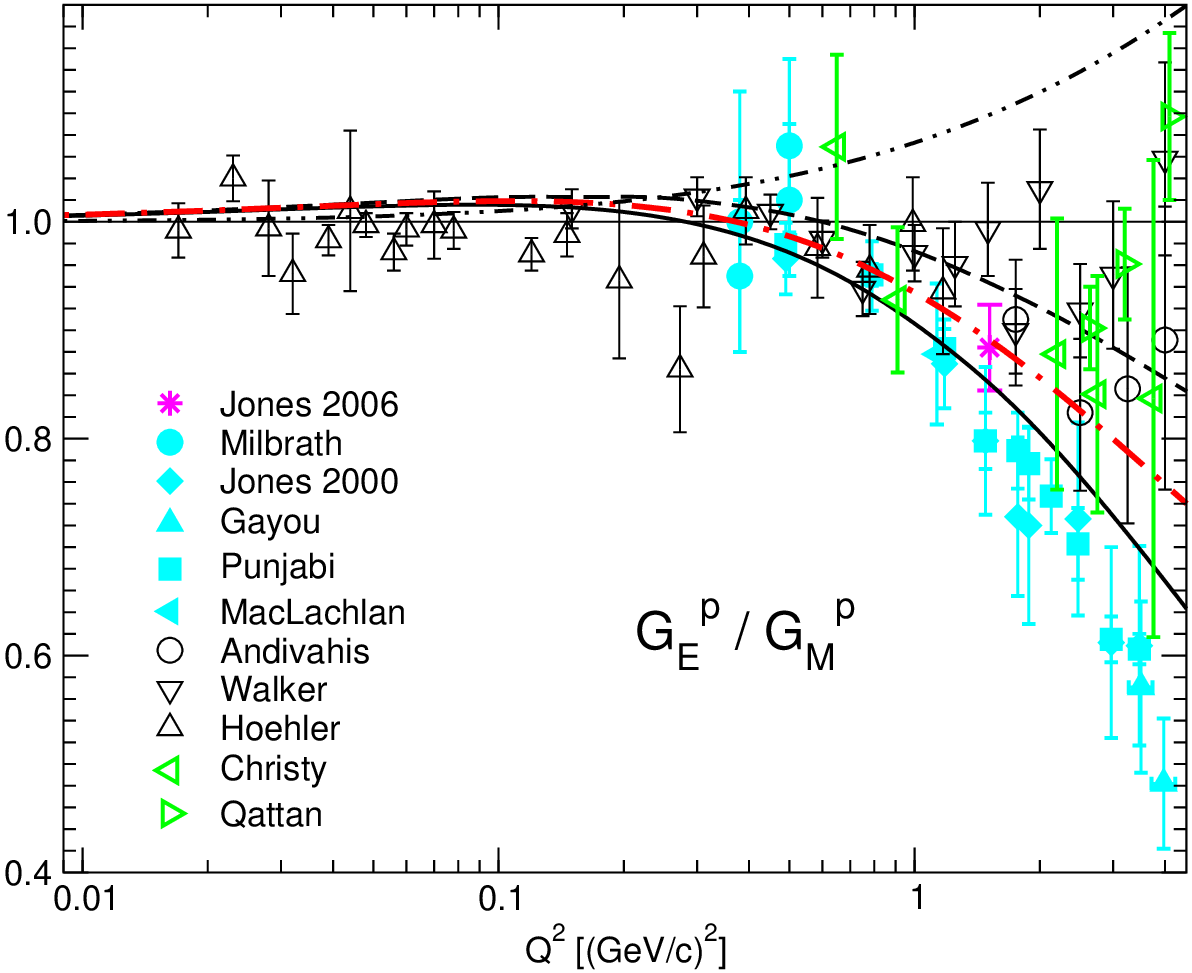}
\caption{\label{fig:ratio_ari_geo_2}
Same comparison as in Fig.~\ref{fig:ratio_ari_geo} but for the
electric to dipole as well as proton electric to magnetic 
form factor ratios. Experimental data from 
Refs.~\protect\cite{%
Andivahis:1994rq,Walker:1989af,Sill:1993qw,Hohler:1976ax,Bartel:1973rf,%
Christy:2004rc,Qattan:2004ht, %
Jones:2006kf,Milbrath:1998de,Jones:1999rz,Gayou:2001qd,Punjabi:2005wq,%
MacLachlan:2006vw%
}. For a better discrimination of the experimental data the values extracted
with the Rosenbluth separation method are marked by open black and green
symbols whereas the direct measurements of the electric to magnetic form-factor
ratio (in the right panel) are depicted by the filled blue and magenta symbols.
}
\end{center}
\end{figure*}

In summary we note that under the given premises there remains a
certain ambiguity in the PFSM construction
related to the factor $\cal N$. Let us thus examine the variations
in the predictions for EM form factors resulting from different possible choices
of $\cal N$.
In Fig.~\ref{fig:ratio_ari_geo} we show the ratios of neutron and proton magnetic
form factors to the standard dipole form factors as obtained in the PFSM with
three particular choices of $\cal N$; the solid line is the ${\cal N}_S$ result and the dashed
line represents the results with ${\cal N}_{\rm ari}$. Fig.~\ref{fig:ratio_ari_geo_2} contains the
same comparisons for the proton electric to magnetic as well as electric to dipole
form factor ratios. The band of variations
in the predictions is limited by the cases with ${\cal N}_{\rm ari}$ and
${\cal N}_{\rm S}$. The uncertainty bands are anyway rather narrow since one
must take into account that these ratios are extremely sensitive to small differences.
For comparison we have added in Figs.~\ref{fig:ratio_ari_geo} and~\ref{fig:ratio_ari_geo_2} 
also the results obtained in IFSM (dashed-double-dotted lines). 

Different choices of ${\cal N}\left(z\right)$ in Eq.~(\ref{eq:facgen}) imply 
distinct $Q^2$-dependences in the form factors. One may ask which particular value of $z$
is favoured by phenomenology.
We have thus performed a simple one-parameter fit of ${\cal N}\left(z\right)$ 
to the experimental data of the ratios $G_M^n/G_D$ and
$G_M^p/G_D$ at a single intermediate momentum transfer of $Q^2=3$~GeV$^2$.
It leads to $z=\frac{1}{6}$ and produces the results represented by
the dash-dotted lines in Fig.~\ref{fig:ratio_ari_geo}. They lie everywhere
in between the upper and lower bounds obtained with ${\cal N}_{\rm ari}$
and ${\cal N}_{\rm S}$, respectively. 

Evidently at $Q^2=0$ the spread from different forms of $\cal N$ vanishes.
It grows towards higher momentum transfers. However, the ambiguity 
band remains relatively small up to $Q^2\approx 3-4$~GeV$^2$.
Regarding the proton form factor ratio in Fig.~\ref{fig:ratio_ari_geo}, the theoretical 
uncertainties are larger than the experimental ones and this represents a significant 
limitation. In addition, one must observe that the results with ${\cal N}_{\rm S}$ 
tend to be lower than the experimental data, at least beyond a momentum transfer
of $Q^2\approx 1$GeV$^2$. However, this curve should not be considered as the most 
favourable PFSM prediction but merely a lower limit of possible results due to 
an intrinsic theoretical uncertainty.
A similar situation replicates also for the ratio of the neutron magnetic to dipole 
form factors in the right panel of Fig.~\ref{fig:ratio_ari_geo}, although in this case the 
experimental uncertainties are larger than for the proton.

The ratios involving the electric proton form factors are given in Fig.~\ref{fig:ratio_ari_geo_2}.
In case of the proton electric to dipole from factor ratio the theoretical uncertainty band from
the PFSM calculations essentially covers the experimental data with their errors.
In particular the PFSM result produced with ${\cal N}_{\rm fit}$, where the $z=\frac{1}{6}$
was adjusted only to the proton and neutron magnetic form factors, is found in
remarkable consistency with the general trend of the experimental data. The IFSM
calculation is not in the position to produce the right $Q^2$-dependence, and it is far off
the established phenomenology.
In the right panel of Fig.~\ref{fig:ratio_ari_geo_2} the ratio of 
the proton electric to magnetic form factors is shown. It has become directly accessible
by experiment through polarization measurements (depicted by the filled blue and 
magenta symbols)
and this has lead to a conflict with earlier cross section data (marked by the open
black symbols); more recently cross-section measurements have produced the data
marked by the open green symbols. It is interesting to observe  that the PFSM predictions
tend to follow the downbending of the ratio with increasing momentum transfer.
The ${\cal N}_{\rm S}$ results agree with the polarization data, the curve with 
${\cal N}_{\rm fit}$ hits the latest measurement reported in Ref.~\cite{Jones:2006kf}
(shown by the magenta star in the right panel of Fig.~\ref{fig:ratio_ari_geo_2}).
The theoretical uncertainty  band, however, remains rather narrow and is in any case smaller
than the spread from the various experimental data.
\section{Summary and Conclusion
\label{sec:Conclusion}
}
We have studied the spectator-model constructions of the electromagnetic
current operator in the instant and point forms of relativistic quantum
mechanics. In particular we have specified the IFSM and PFSM current operators
and have derived their common nonrelativistic limit, the usual NRIA. We have
calculated the direct predictions of the GBE CQM for the elastic nucleon EM form
factors, including the electric radii and magnetic moments. The PFSM results are
found close to experimental data in all instances, while the IFSM and NRIA
results deviate grossly.

Furthermore, we have investigated the dependences of the form-factor
results on nucleon wave functions from different relativistic CQMs.
The predictions of the GBE CQM in PFSM have been contrasted to analogous results
calculated with the relativistic BCN CQM. The two CQMs rely on hyperfine interactions 
from rather distinct dynamical concepts
(flavour dependent Goldstone-boson exchange and color-magnetic interactions,
respectively). Nevertheless, only minor variations are found in all predictions 
for elastic nucleon form factors (as well as electric radii and magnetic moments).
In addition we have provided a comparison with the predictions of the II CQM,
derived within a completely different approach along the Bethe-Salpeter formalism.
Still, the II CQM leads to results that are quite similar 
to the PFSM ones, in spite of the completely different frameworks.
Instead, if one leaves out the hyperfine interaction completely, one obtains an
unrealistic description that leads specifically to an almost vanishing electric
form factor of the neutron.

In addition we have addressed a theoretical uncertainty that resides 
in the construction of the PFSM current. It concerns the choice of a 
factor $\cal N$ in the PFSM construction that is unavoidable and cannot be
constrained uniquely by Poincar\'e invariance alone. Supplementary conditions
such as charge normalization and time-reversal invariance can be imposed.
Nevertheless, there remains a residual indetermination in the factor $\cal N$.
We demonstrated the band of variations of the PFSM results due to different
possible choices for $\cal N$. It typically covers the spread of experimental
data with the upper bound represented by ${\cal N}_{\rm ari}$ and the
lower one by ${\cal N}_{\rm S}$. The magnitudes of the variations between
results with different factors of $\cal N$, however, are generally larger
than differences due to nucleon wave functions from alternative quark models.

At present there is a vivid discussion of some details of the proton
electromagnetic structure revealed by experiment through the $G_E^p/G_M^p$
ratio as depicted in Fig.~\ref{fig:ratio_ari_geo_2}. Specifically one has
become aware of a striking discrepancy between earlier data extracted by
the Rosenbluth separation
method~\cite{Andivahis:1994rq,Walker:1989af,Hohler:1976ax} and
more recent data measured in polarization 
experiments~\cite{Jones:1999rz,Gayou:2001qd,Punjabi:2005wq,MacLachlan:2006vw}.
The problem has recently been investigated experimentally 
by additional measurements at the Jefferson Laboratory using the
Rosenbluth technique~\cite{Christy:2004rc,Qattan:2004ht}.
According to Ref.~\cite{Arrington:2004is} the inclusion of Coulomb distortions
in the Rosenbluth method has a non-negligible effect, but cannot account for
the whole discrepancy. It also has been suggested that the effect of two-photon
contributions could have an impact~\cite{Arrington:2003qk,Arrington:2004ae},
particularly in the Rosenbluth separation. Obviously, the experimental situation
is still a matter of discussion (see 
also Refs.~\cite{Tvaskis:2005ex,Jones:2006kf,Tomasi-Gustafsson:2006pa}).
On the theoretical side, the major efforts to resolve the discrepancies
take into account two-photon 
corrections~\cite{Guichon:2003qm,Blunden:2003sp,Blunden:2005ew,%
Chen:2004tw,Afanasev:2005mp}
and additional $\Delta$ contributions~\cite{Kondratyuk:2005kk}. For an updated
discussion on this issue see, e.g., Ref.~\cite{Arrington:2006zm}.

Our present study tells that the IFSM calculation is not in the position
to reproduce the $G_E^p/G_M^p$ data from neither the Rosenbluth separation nor
from the polarization measurements. The corresponding momentum dependence is
contrary to both. On the other hand, one might be tempted to conclude
that the usual PFSM calculation (with the factor $\cal N_{\rm S}$) favours
the lower lying polarization data (cf. Fig.~\ref{fig:ratio_ari_geo_2}). However,
our investigation of the uncertainties still inherent in the PFSM
results tells that the upper bound of the predictions also comes close to
the data from Rosenbluth separation. With regard to the recent datum from the 
asymmetric beam-target experiment by Jones et al.~\cite{Jones:2006kf} it is
particularly interesting to see that it is hit by the prediction with
$\cal N_{\rm fit}$. 

\begin{acknowledgments}
This work was supported by the Austrian Science Fund 
(Projects P16945 and  P19035)
and the Italian MIUR-PRIN Project ``Struttura nucleare e reazioni nucleari dai 
pochi corpi ai molti corpi". T.M  and W.P. are grateful to the INFN, Sezione di Padova, and 
the University of Padova for supporting their visits. 
L.C. thanks the University of Graz for the hospitality extended to him. 
The authors acknowledge useful discussions with A. Krassnigg and W. Schweiger.
\end{acknowledgments}
\end{document}